\documentclass[]{aa}
\usepackage[utf8]{inputenc}

\usepackage{graphicx}
\graphicspath{{./}}
\usepackage{amssymb,amsmath,amsfonts}
\usepackage{txfonts}

\usepackage{epsfig}
\usepackage{color}
\usepackage[dvipsnames]{xcolor}
\usepackage{url}
\usepackage[]{hyperref}
\hypersetup{colorlinks=true,linkcolor=blue,citecolor=blue,filecolor=blue,
urlcolor=blue}

\usepackage{natbib}
\bibpunct{(}{)}{;}{a}{}{,}

\begin{document}

   \title{The LoTSS view of radio AGN in the local Universe}
   \subtitle{The most massive galaxies are always switched on}
   \titlerunning{The LoTSS view of radio AGN in the local Universe}
   \author{J.~Sabater\inst{1}\thanks{E-mail: jsm@roe.ac.uk}
          \and
          P.~N.~Best\inst{1}
          \and 
          M.~J.~Hardcastle\inst{2}
          \and 
          T.~W.~Shimwell\inst{3}
          \and 
          C.~Tasse\inst{4,}\inst{5}
          \and 
          W.~L.~Williams\inst{2}
          \and
          M.~Br\"uggen\inst{6}
          \and
          R.~K.~Cochrane\inst{1}
          \and 
          J.~H.~Croston\inst{7}
          \and 
          F.~de~Gasperin\inst{6}
          \and 
          K.~J.~Duncan\inst{8}
          \and
          G.~G\"urkan\inst{9}
          \and
          A.~P.~Mechev\inst{8}
          \and
          L.~K.~Morabito\inst{10}
          \and 
          I.~Prandoni\inst{11}
          \and 
          H.~J.~A.~R\"ottgering\inst{8}
          \and 
          D.~J.~B.~Smith\inst{2}
          \and 
          J.~J.~Harwood\inst{2}
          \and
          B.~Mingo\inst{7}
          \and 
          S.~Mooney\inst{12}
          \and 
          A.~Saxena\inst{8}
          }

    \institute{
SUPA, Institute for Astronomy, Royal Observatory, Blackford Hill, 
Edinburgh, EH9 3HJ, UK
\and
Centre for Astrophysics Research, School of Physics, Astronomy and 
Mathematics, University of Hertfordshire, College Lane, Hatfield AL10 9AB, UK
\and
ASTRON, the Netherlands Institute for Radio Astronomy, Postbus 2, 7990 
AA, Dwingeloo, The Netherlands
\and
GEPI, Observatoire de Paris, CNRS, Universite Paris Diderot, 5 place 
Jules 
Janssen, 92190 Meudon, France
\and
Department of Physics \& Electronics, Rhodes University, PO Box 94, 
Grahamstown, 6140, South Africa
\and
University of Hamburg, Hamburger Sternwarte, Gojenbergsweg 112, 21029 
Hamburg, Germany
\and
School of Physical Sciences, The Open University, Walton Hall, Milton 
Keynes, MK7 6AA, UK
\and
Leiden Observatory, Leiden University, PO Box 9513, NL-2300 RA Leiden, 
The Netherlands
\and
CSIRO Astronomy and Space Science, PO Box 1130, Bentley WA 6102, 
Australia
\and
Astrophysics, University of Oxford, Denys Wilkinson Building, Keble Road, 
Oxford, OX1 3RH, UK
\and
INAF -- Istituto di Radioastronomia, via Gobetti 101, 40129 Bologna, 
Italy
\and
School of Physics, University College Dublin, Belfield, Dublin 4, 
Republic of Ireland
             }

   \date{Accepted ---; received ---; in original form \today}
   
   \abstract
{
This paper presents a study of the local radio source population, by 
cross-comparing the data from the first data release (DR1) of the LOFAR 
Two-Metre Sky Survey (LoTSS) with the Sloan Digital Sky Survey (SDSS) DR7 main 
galaxy spectroscopic sample. The LoTSS DR1 provides deep data (median rms noise 
of 71\,$\mathrm{\mu}$Jy at 150\,MHz) over 424 square degrees of sky, which is 
sufficient to detect 10615 (32 per cent) of the SDSS galaxies over this sky 
area. An improved method to separate active galactic nuclei (AGN) accurately 
from sources with radio emission powered by star formation (SF) is developed and 
applied, leading to a sample of 2121 local ($z < 0.3$) radio AGN. The local 
150\,MHz luminosity function is derived for radio AGN and SF galaxies 
separately, and the good agreement with previous studies at 1.4\,GHz suggests 
that the separation method presented is robust. The prevalence of radio AGN 
activity is confirmed to show a strong dependence on both stellar and black hole 
masses, remarkably reaching a fraction of 100 per cent of the most massive 
galaxies ($> 10^{11} \mathrm{M_{\odot}}$) displaying radio-AGN activity with 
$L_{\rm 150\,MHz} \geq 10^{21}$W\,Hz$^{-1}$; thus, the most massive galaxies are 
always switched on at some level. The results allow the full Eddington-scaled 
accretion rate distribution (a proxy for the duty cycle) to be probed for 
massive galaxies, and this accretion rate is found to peak at 
$L_{\mathrm{mech}}/L_{\mathrm{Edd}} \approx 10^{-5}$. More than 50 per cent of 
the energy is released during the $\le 2$ per cent of the time spent at the 
highest accretion rates, $L_{\mathrm{mech}}/L_{\mathrm{Edd}} > 10^{-2.5}$. 
Stellar mass is shown to be a more important driver of radio-AGN activity than 
black hole mass, suggesting a possible connection between the fuelling gas and 
the surrounding halo. This result is in line with models in which these radio 
AGN are essential for maintaining the quenched state of galaxies at the centres 
of hot gas haloes.
}

\keywords{surveys -- galaxies: evolution -- galaxies: active -- radio 
continuum: galaxies}
               
\maketitle


\section{Introduction}
\label{sec:intro}

The large-scale structure of the Universe is driven by the gravitational 
collapse of dark matter haloes and the subsequent merging of these into 
progressively larger structures (filaments, sheets, groups, and clusters) 
building the ``cosmic web''. Galaxies then form within dark matter haloes by 
radiative cooling of baryons \citep{White1978}. The efficiency by which baryons 
are converted into stars is highly dependent on the mass of the dark matter 
halo: the stellar-mass to halo-mass relation peaks at $M_{\rm halo} \approx 
10^{12} \mathrm{M_{\odot}}$, at which mass around 20 per cent of baryons are 
converted into stars \citep[e.g.][]{Behroozi2010,Moster2010,Wechsler2018}.

At halo masses below this peak, the heating of gas by ultraviolet radiation from 
both the host galaxy and cosmic background, and the ejection of gas by 
supernova-driven winds \citep[e.g.][]{Benson2003,Baugh2006}, combine to lower 
the SF efficiency in galaxies. At halo masses above $10^{12} 
\mathrm{M_{\odot}}$, the overall SF efficiency of haloes is reduced from SF 
being terminated, or quenched, in a subset of the galaxy population.  
\citet{Peng2010} studied the fraction of passive, red galaxies in the local 
Universe as a function of both mass and environment and argued that there were 
two distinct mechanisms for this quenching of SF, which operates independently: 
`environment quenching', which applies mainly to satellite galaxies, and `mass 
quenching' of the most massive galaxies. Subsequent work has broadly confirmed 
these trends \citep{Peng2012,Gabor2015,Zu2016,Cochrane2018}. Nevertheless, the 
physical drivers of these quenching mechanisms remain widely debated, in 
particular the extent to which mass quenching is driven by internal galactic 
processes such as active galactic nuclei (AGN) as compared to the role of the 
dark matter halo.

Mass quenching has frequently been attributed to the effects of AGN because AGN 
activity is observed to occur in the high-mass galaxies in which SF needs to be 
quenched \citep[e.g.][]{Heckman2014} and because it is well established 
\citep[e.g.][]{Silk1998,Fabian1999,King2003,Fabian2012} that winds and outflows 
driven by quasar activity can terminate SF in a manner that also gives rise to 
the observed correlation between the mass (or velocity dispersion) of a galaxy 
bulge and the mass of the central black hole (e.g.\ 
\citealt{Magorrian1998,Gebhardt2000}; see review by \citealt{Kormendy2013}). 
However, mass quenching could also be closely linked to halo properties owing to 
the strong correlation between stellar mass and halo mass. As noted by 
\citet{Bower2006}, when gas falls into a dark matter halo of mass below a few 
$\times 10^{11} \mathrm{M_{\odot}}$, the timescale for it to cool is shorter 
than its infall time, and so it arrives at the central galaxy cold and is able 
to be efficiently converted into stars. At higher halo masses, however, the 
infalling gas suffers a virial shock and its cooling time exceeds the dynamical 
time, leading to a hydrostatic halo of hot (X-ray emitting) gas that is built up 
\citep[e.g.][]{Birnboim2003,Keres2005}. This transition in the nature of 
accreted gas leads to a natural explanation for the decrease in SF efficiency in 
high-mass haloes.

Regardless of the initial quenching mechanism, the hot gas in massive haloes 
radiates and cools and, especially in rich, undisturbed environments, might be 
expected to form a cooling flow \citep[see][]{Fabian1994}, which would result in 
high levels of gas deposition and star formation (SF) in the central galaxy. A 
source of heating is required to offset the cooling and prevent this. It is now 
widely accepted that radio AGN are responsible for this: radio-AGN activity 
(either current or recently terminated) is seen in essentially all of the 
central galaxies of cool-core clusters \citep{Burns1990,Dunn2006,Best2007} and 
these AGN are able to deposit the jet energy directly and efficiently into the 
intracluster medium by inflating bubbles/cavities \citep[see reviews 
by][]{McNamara2007,Fabian2012}. Estimates of the radio-AGN heating rate in these 
clusters \citep[as determined from the cavity enthalpy coupled with a buoyancy 
timescale; e.g.][]{Birzan2004,Cavagnolo2010} show that these are well matched to 
the radiative cooling rates of the hot gas \citep[e.g.][]{McNamara2007}. This 
forms a natural feedback cycle, whereby the hot gas offers both a source of fuel 
for the radio AGN through cold chaotic accretion \citep[e.g.][]{Gaspari2013} and 
a confining medium for the radio source to expand against and deposit its energy 
into \citep[see review by][]{Heckman2014}.

Radio AGN feedback is likely to be important on galactic scales as well and 
massive galaxies are also located in hot hydrostatic envelopes 
\citep[e.g.][]{Croston2007, Mingo2011, Mingo2012}. It was the inclusion of this 
`jet-mode' (or `maintenance-mode') feedback from radio AGN into semi-analytic 
models of galaxy formation that allowed \citet{Croton2006} and \citet{Bower2006} 
to naturally explain the shape of the galaxy luminosity function and the bimodal 
nature of the galaxy population. Modern hydrodynamical simulations, including a 
feedback term (associated with radio AGN) that suppresses gas cooling in hot 
haloes, are also successful at reproducing the local observed trends in galaxy 
properties \citep[e.g.][]{Gabor2015}. These results provide support for 
so-called halo-quenching models for the switch off of SF in massive galaxies.

Observationally, considerable advances in our understanding of the importance of 
radio AGN have been made over the last one to two decades; a major driver of 
this has been the availability of uniform, wide-area spectroscopic surveys such 
as the Two-degree Field Galaxy Redshift Survey \citep[2dFGRS;][]{Colless2001} 
and the Sloan Digital Sky Survey \citep[SDSS;][]{York2000,Stoughton2002}, 
coupled with wide-area radio surveys, especially the National Radio Astronomy 
Observatory (NRAO) Very Large Array (VLA) Sky Survey \citep[NVSS;][]{Condon1998} 
and the Faint Images of the Radio Sky at Twenty centimetres survey 
\citep[FIRST;][]{Becker1995}. Cross-matching of radio and optical surveys 
\citep[e.g.][]{Sadler2002,Best2005a,Mauch2007,Best2012} has allowed detailed 
statistical studies of the prevalence and properties of radio-AGN activity.

\citet{Best2005b} showed that the fraction of massive galaxies that host 
radio-AGN activity is a very strong function of stellar mass ($f_{\rm rad} 
\propto M_*^{2.5}$) or black hole mass ($f_{\rm rad} \propto 
M_{\mathrm{BH}}^{1.6}$), reaching as high as $\approx 30$ per cent at the 
highest stellar masses, to the radio luminosity limit of their analysis 
($L_{\mathrm{1.4\,GHz}} \approx 10^{23}$W\,Hz$^{-1}$); see also \cite{Brown2011} 
for a deeper study of a much smaller sample. \citet{Best2006} and 
\citet{Best2007} built upon this to estimate the time-averaged heating rate 
associated with this radio-AGN activity, assuming that all massive galaxies go 
through recurrent radio-AGN outbursts and that the observed radio-AGN prevalence 
could be used as a measure of the duty cycle of the AGN activity. They found 
that this heating rate exceeded the heating rate that is necessary to 
counterbalance the typical radiative cooling losses of the hot gas, and that 
radio-AGN heating is therefore able to explain the old, red, and dead nature of 
massive galaxies in hot haloes.

Although this broad understanding is in place, several open questions remain 
about the detailed process of radio-AGN feedback. One such question concerns the 
duty cycle of radio-AGN activity. \citet{Kauffmann2009} and \citet{Best2012} 
studied the distribution of Eddington-scaled accretion ratios at low Eddington 
ratios and found that, down to their limit of around $L/L_{\rm Edd} = 10^{-3}$, 
these follow roughly a power law with increasing numbers of low Eddington ratio 
sources \citep[see also][]{Heckman2014}. Deeper radio data are needed to probe 
lower radio luminosities, and hence track the full distribution of accretion 
rates, over wide enough sky areas to build significant samples. Such data would 
also allow a much larger fraction of lower stellar mass sources to be detected, 
enabling investigation of the relationships between duty cycle, radio luminosity 
distribution, and stellar mass. 

The LOFAR Two-Metre Sky Survey \citep[LoTSS; ][]{Shimwell2017} offers a new 
opportunity to advance these studies and address some of these questions. This 
large-area 150\,MHz survey with the LOw Frequency ARray \citep[LOFAR; 
][]{vanHaarlem2013} reaches more than an order of magnitude deeper than the 
FIRST survey for sources of typical spectral index, and is even more 
advantageous for steep spectrum sources (older electron populations). 
Furthermore, it also has high sensitivity to extended radio structure, thus 
avoiding the need for the complicated combination of FIRST and NVSS that 
\citet{Best2005a} needed to adopt. This paper cross-matches data from the first 
data release \citep[DR1; ][]{Shimwell2018} of LoTSS with the main galaxy sample 
of the SDSS to provide new insights into the prevalence, duty cycle, and impact 
of radio-AGN activity in the local Universe.

The layout of the paper is as follows. Section~\ref{sec:sample_data} describes 
the sample of galaxies to be studied and the cross-matching between the radio 
and optical catalogues. Section~\ref{sec:radio_loud} then outlines the procedure 
to filter the radio-AGN subset of these sources from the bulk population of 
star-forming galaxies (SFGs). The radio spectral index properties of these 
sources, and the resultant local radio luminosity function at 150\,MHz are 
discussed in Sections~\ref{sec:spectral_index} and~\ref{sec:lum_func}, 
respectively. Section~\ref{sec:fractions} then considers the prevalence of radio 
AGN as a function of both stellar mass and black hole mass, comparing the 
results to previous studies, and using the large sample to break the degeneracy 
between these two parameters. Section~\ref{sec:accretion_rates} examines the 
distribution of Eddington-scaled accretion rates of the most massive galaxies. 
Finally, the results of the paper are summarised in 
Section~\ref{sec:conclusions}, and their implications are discussed. Throughout 
the study, a cosmology with $\mathrm{\Omega}_{\mathrm{\Lambda0}} = 0.7$, 
$\mathrm{\Omega}_{\mathrm{m0}} = 0.3$, and $\mathrm{H}_{\mathrm{0}} = 70$ km 
s$^{-1}$ Mpc$^{-1}$ is assumed.


\section{The sample and data}
\label{sec:sample_data}

The LoTSS \citep[][]{Shimwell2017} is a high-resolution survey that will cover 
the full northern hemisphere at frequencies ranging from 120 to 168 
MHz\footnote{Although the central frequency of the LoTSS band is 144\,MHz, the 
bandwidth is large and the sensitivity-weighted mean frequency depends on 
position within the mosaic owing to the frequency-dependent primary beam size. 
To avoid an impression of undue precision, we use 150\,MHz throughout the paper 
to refer to the LoTSS frequency.}. The LoTSS DR1 \citep[][]{Shimwell2018} covers 
424 square degrees centred in the Hobby-Eberly Telescope Dark Energy Experiment 
\citep[HETDEX; ][]{Hill2008} Spring Field region (right ascension 10h45m00s to 
15h30m00s and declination 45$^\circ$00$\arcmin$00$\arcsec$ to 
57$^\circ$00$\arcmin$00$\arcsec$) and contains over 300,000 sources with a 
signal of at least five times the noise level. It is composed of 58 overlapping 
pointings with 8 hours of observation in each one. The median rms noise is 
$\approx$ 71\,$\mathrm{\mu}$Jy~beam$^{-1}$ and 95 per cent of the area in the 
release has an rms noise level below 150\,$\mathrm{\mu}$Jy~beam$^{-1}$. The 
angular resolution is 6 arcseconds and the positional accuracy is better than 
0.2 arcseconds for high signal-to-noise sources; the positional accuracy 
increases to $\approx$ 0.5 arcseconds for the faintest sources with a flux 
density of less than 0.6 mJy. Detailed information about the LoTSS DR1 is 
presented in \citet{Shimwell2018}. The radio sources were associated with their 
optical and mid-infrared counterparts in the Panoramic Survey Response and Rapid 
Response System \citep[Pan-STARRS;][]{Kaiser2002, Kaiser2010} and the Wide-field 
Infrared Survey Explorer \citep[WISE;][]{Wright2010} surveys (with a 
completeness and reliability at the $\approx 99$ per cent level) using a 
combination of statistical likelihood ratio techniques and visual classification 
methods \citep{Williams2018}. Finally, photometric redshifts were derived and 
rest-frame magnitudes were obtained for the sources \citep[][although those data 
are not used in the current paper]{Duncan2018}.

The galaxy sample selected for this study is based on the spectroscopic data 
from the seventh data release of the SDSS  \citep[SDSS DR7;][]{Abazajian2009}. 
There are 34709 galaxies from the SDSS Main Galaxy Sample \citep[broadly 
complete between magnitudes 14.5 and 17.77 in r band;][]{Strauss2002} that 
overlap with the region covered by the LoTSS DR1. An upper redshift limit of $z 
= 0.3$ is applied, which reflects the approximate limit of main galaxy sample 
spectroscopy ($n_{z > 0.3} = 47$). A lower redshift limit of $z=0.01$ is also 
applied, as below this the SDSS fibres probe such a small region of the galaxy 
that derived galaxy parameters become unreliable ($n_{z < 0.01} = 413$). 
Galaxies with a non-zero $z$ warning flag were discarded as their redshifts 
could be unreliable ($n_{\mathrm{z\,warning}} = 34$). The catalogue contains 
duplicated observations for some of the objects and a single observation of each 
object was selected to avoid multiple counting on the statistics 
($n_{\mathrm{duplicates}} = 711$).  With all of these constraints, a sample of 
33504 galaxies was obtained in the LoTSS DR1 area.

All the SDSS sources are bright in the optical band and have clear corresponding 
PanSTARRS detections. Therefore, the radio counterparts of SDSS galaxies were 
found by cross-matching the positions of SDSS galaxies with the PanSTARRS 
counterparts of the LoTSS sources. LoTSS sources present accurate 
PanSTARRS-LoTSS cross identifications even for extended sources due to the 
efforts shown in \citet{Williams2018}. The positions of the SDSS sources were 
cross-matched with the optical (PanSTARRS) positions using a search radius of 2 
arcseconds. This radius is sufficiently large to obtain essentially all the 
genuine counterparts with no significant contamination by random matches. In any 
case, the results are largely insensitive to this choice of cross-match radius 
if it is within the range of 1 to 2 arcseconds since the PanSTARRS to SDSS 
astrometric agreement is typically accurate to fractions of an arcsecond. A 
LoTSS counterpart was found for 10615 of the SDSS sample (32 per cent). The 
LoTSS data release provides mosaics with the rms noise level for each position, 
which also allows the flux density limit at 150 MHz to be determined for the 
SDSS galaxies without a LoTSS cross-match. This limit was set at five times the 
rms noise level, which corresponds to the approximate catalogue limit for LoTSS 
\citep{Shimwell2018}. There are 43 sources for which the measured LoTSS total 
flux density is slightly below five times the rms noise. For completeness these 
sources were removed from the catalogue and considered as non-detections.

The MPA-JHU\footnote{https://wwwmpa.mpa-garching.mpg.de/SDSS/DR7/} value-added 
catalogue \citep{Brinchmann2004arxiv} offers improved and additional information 
to the SDSS data. In particular, the nebular emission lines were measured after 
the subtraction of the contribution of the stellar populations 
\citep{Tremonti2004} and are used for the selection of AGN in the next section. 
We also use the stellar mass of the galaxies presented in the catalogue, as 
derived by \citet{Kauffmann2003a}. Additional parameters used in this paper are 
the 4000\AA\ break strength ($D_{\rm 4000}$), and the velocity dispersion 
($\sigma$). The latter can be used to estimate the mass of the central black 
hole: the relation adopted in this paper is that of \citet{McConnell2013}, i.e. 
$\log_{10}(M_{\mathrm{BH}}/\mathrm{M_{\odot}}) = 8.32 + 5.64 
\log_{10}[\sigma/(200 \mathrm{km\,s^{-1}})]$. \citet{Best2012} cross-matched the 
SDSS DR7 spectroscopic catalogue against the NVSS and FIRST radio surveys to 
derive the 1.4\,GHz properties of these sources; data from their catalogues are 
also used in the analysis.


\section{Selection of radio AGN}
\label{sec:radio_loud}

A significant challenge in deep radio surveys such as LoTSS is to separate radio 
AGN from SFGs, for which the radio emission is associated with the SF in the 
galaxy, but there may or may not also be a radio-quiet AGN present; radio AGN 
are defined in this work as galaxies for which their radio emission is dominated 
by a jet originating from the central supermassive black hole.  Star-forming 
galaxies emit at radio wavelengths primarily due to synchrotron emission from 
shocks associated with supernovae, and hence their radio luminosity correlates 
broadly with the SFR (e.g.\ \citealt{Condon1992}), although there are 
indications that the relationship can also depend on other parameters such as 
galaxy mass (e.g.\ \citealt{Gurkan2018,Read2018}). Star-forming  galaxies 
dominate the radio source counts at flux densities $S_{\rm 150\,MHz} \le 1\,$mJy 
\citep{Wilman2008,deZotti2010,Williams2016,CalistroRivera2017} and cannot be 
ignored at higher flux densities.

A common method of separating SFGs from AGN is through their optical 
emission-line properties, in particular using emission line diagnostic diagrams 
\citep{Baldwin1981,Kauffmann2003,Kewley2006}. One shortcoming of this method is 
that it cannot be used for objects without clear emission line detections. 
Although for radio-selected samples, these sources are mostly weak-lined radio 
AGN \citep[e.g.][]{Sadler2002}. An additional complication is the radio-quiet 
AGN. The origin of the radio emission in these sources remains widely debated, 
with both SF and weak radio jets contributing, and the dominant mechanism 
varying between sources \citep[e.g.][and references therein]{White2017}. In the 
local Universe, regardless of the presence or not of weak jets, radio-quiet AGN 
are frequently associated with SFGs (e.g.\ \citealt{Kauffmann2003}; see also 
discussion in \citealt{Heckman2014}) and, at the depth of surveys like LoTSS, 
such SF may lead to the galaxy being detected in the radio. The emission line 
fluxes of such sources can be driven by the hard ionising spectrum of the AGN, 
leading to an AGN classification by optical line diagnostics, but such 
SF-dominated sources should not be included in samples of radio AGN.

An alternative approach to identify radio AGN is to compare the radio and 
far-infrared luminosities of sources \citep[e.g.][]{Machalski1999, Sabater2012, 
Mingo2016}. This technique makes use of the relatively tight far-infrared radio 
correlation (FIRC) of SFGs \citep[e.g.][]{Yun2001, Smith2014}: objects with a 
radio excess above that predicted from the FIRC can be identified as radio AGN. 
This is a powerful method (albeit still  limited by the intrinsic scatter in the 
FIRC), but is generally not applicable because of the absence of sufficiently 
deep far-IR data.

\citet{Best2005b} proposed an alternative method for identification of radio 
AGN, using the `$D_{\rm 4000}$ versus $L_{\rm rad}$/$M_*$' plane. The parameter 
$D_{\rm 4000}$ is the strength of the 4000\AA\ break in the galaxy spectrum, and 
$L_{\rm rad}$/$M_*$ is the ratio of radio luminosity to stellar mass. 
\citeauthor{Best2005b} showed that for a wide range of different SF histories, 
SFGs occupy largely the same region of this plane, since both $D_{\rm 4000}$ and 
$L_{\rm rad}$/$M_*$ depend broadly on the specific SFR of the galaxy. 
\citeauthor{Best2005b} used these theoretical tracks to propose a diagnostic 
division line between radio AGN and SFGs whereby the radio-quiet AGN were 
classified together with the SFGs.

\citet{Kauffmann2008} considered an alternative diagnostic, making use of the 
ratio of radio luminosity to H$\alpha$ luminosity. In this work, the latter is 
used as a proxy for SFR, following the same principles as the separation via the 
FIRC. Using this separation they also proposed a modification to the division 
line on the $D_{\rm 4000}$ versus $L_{\rm rad}$/$M_*$ diagnostic of 
\citet{Best2005b}. \citet{Best2012} then further built upon this work by 
devising a scheme to separate radio AGN from SFGs using a combination of the 
revised $D_{\rm 4000}$ versus $L_{\rm rad}$/$M_*$ diagnostic, the emission line 
diagnostic diagram, and the radio luminosity to H$\alpha$ luminosity relation.

In this paper, the identification of the radio-AGN subsample builds upon the 
work of \citet{Best2012}, but with some significant modifications. First, the 
division lines are recast into 150\,MHz luminosities (as detailed in the next 
subsection) owing to the observing frequency of LoTSS. Second, a fourth 
diagnostic test is also included in the analysis, made feasible by the 
availability of WISE data.  Star-forming galaxies separate from the typical 
hosts of radio AGN in their WISE colours, particularly in W2--W3 \citep[4.6 to 
12 micron colour;][]{Yan2013}: \citet{Herpich2016} use a cut at W2--W3 = 2.5 (in 
Vega magnitudes; W2--W3 $\approx 0.7$ in AB magnitudes) to separate galaxies 
with and without SF, although it is also clear from their plots that this 
separation is far from clean. Third, we make use of the analysis by 
\citet{Gurkan2018} of the LOFAR data in the Herschel-ATLAS North Galactic Plane 
(hereafter H-ATLAS) field \citep{Hardcastle2016}. These LOFAR data are similar 
to those of LoTSS and have the same SDSS and WISE properties available. But 
\citet{Gurkan2018} have done detailed spectral energy distribution fitting, 
incorporating the available Herschel data in a self-consistent manner using 
\textsc{MAGPHYS} (\citealt{daCunha2008}; see also \citealt{Smith2012}) to derive 
accurate SFRs and allow more precise separation of SFGs from radio AGN based on 
the radio luminosity to SFR relation. This complementary sample therefore 
provides an opportunity to test and optimise the calibration of the diagnostic 
division lines, and the resultant combined classification scheme, as outlined 
below.

\subsection{Adopted individual diagnostics for population separation}
\label{sec:sfagndiags}

\begin{figure*}
  \begin{tabular}{cc}
    \psfig{file=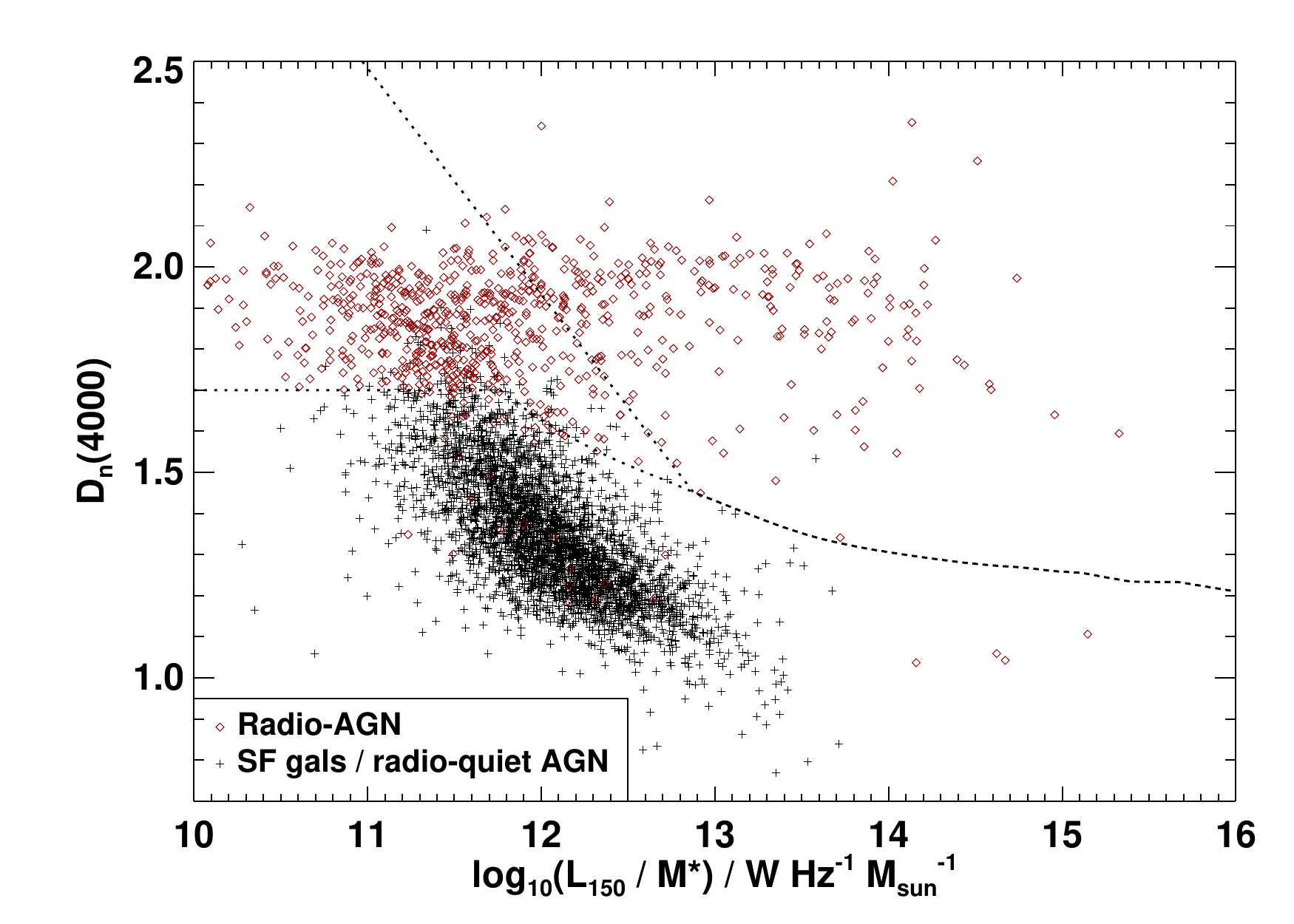,width=8.8cm,clip=} &
    \psfig{file=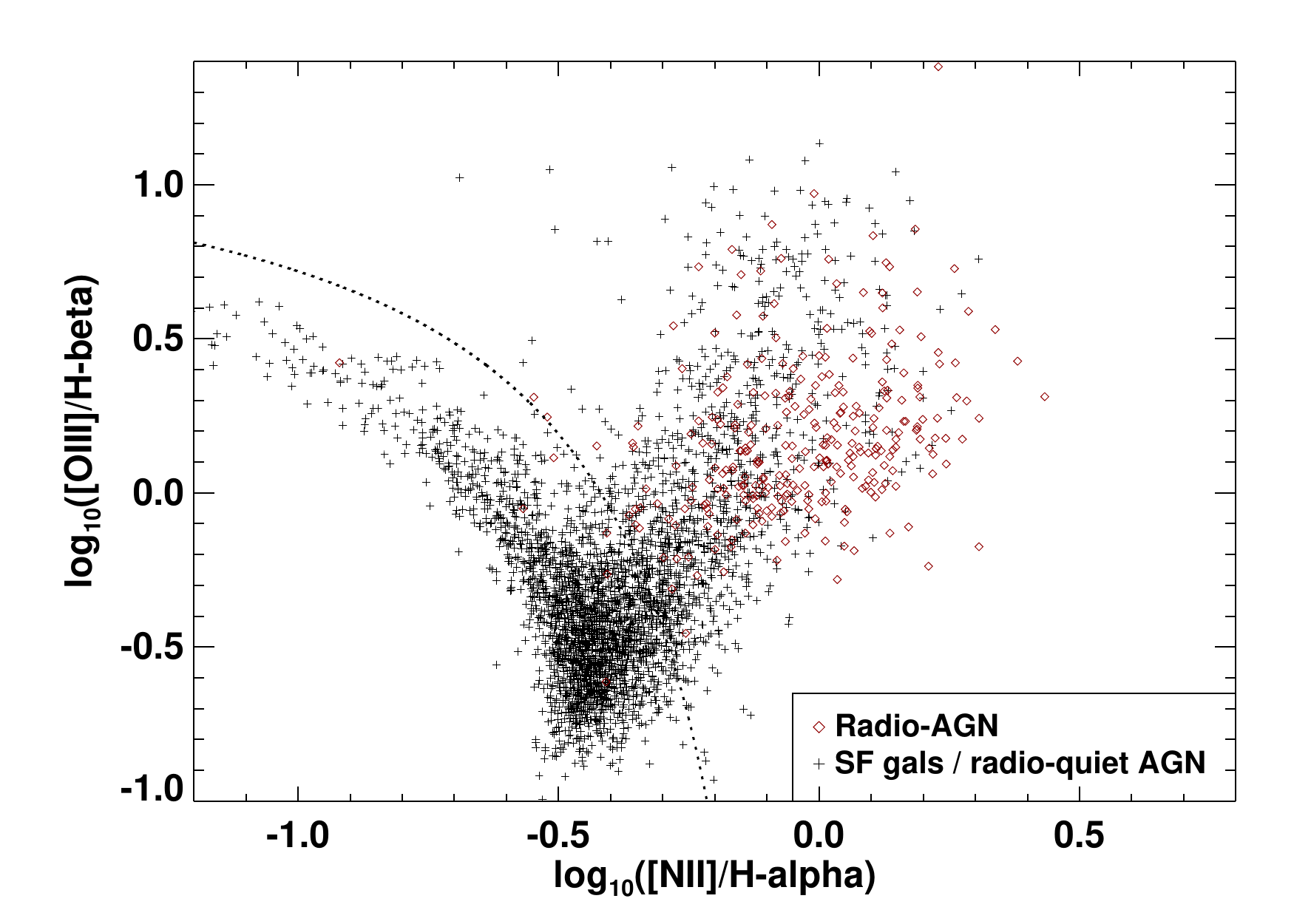,width=8.8cm,clip=} \\
    \psfig{file=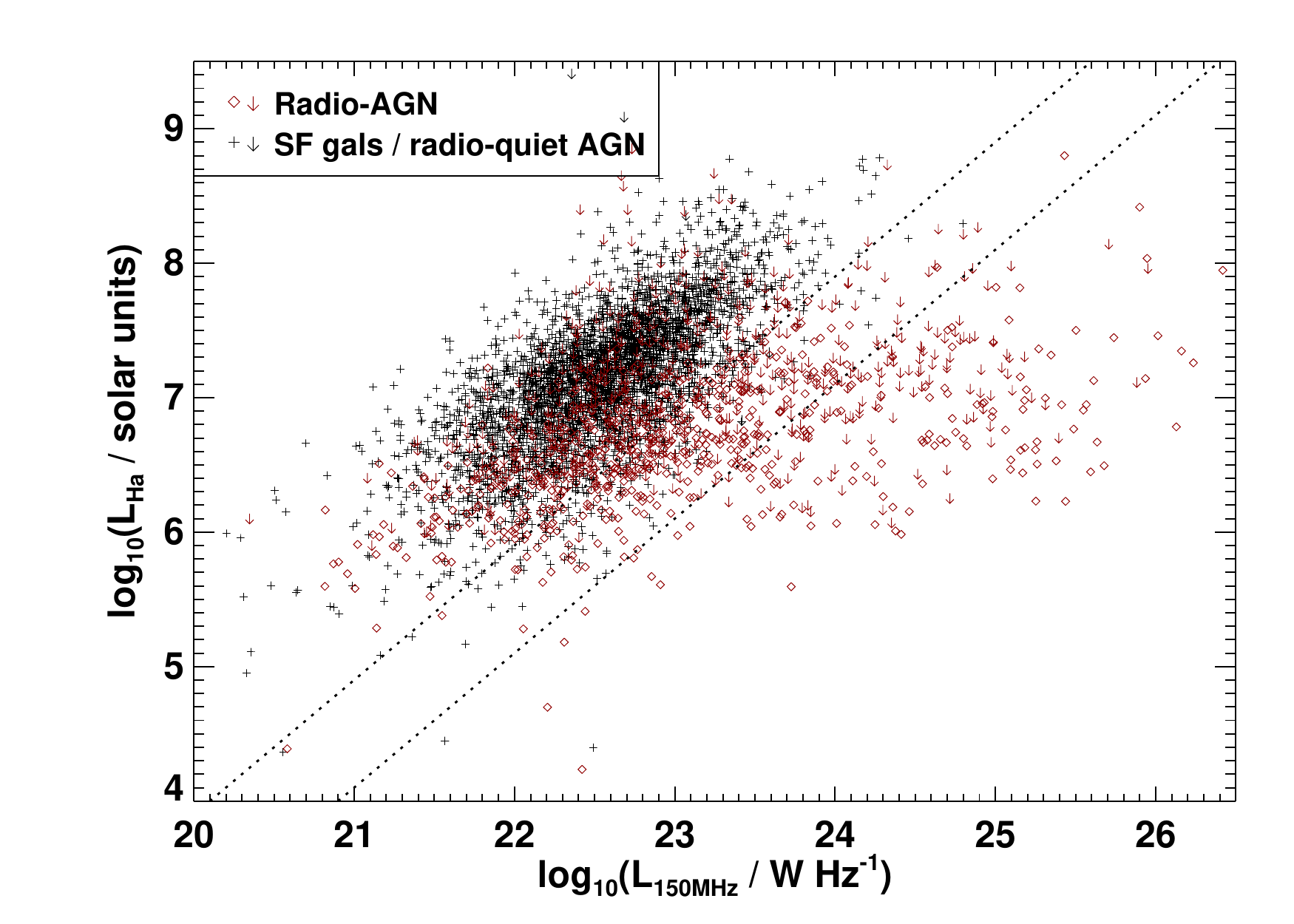,width=8.8cm,clip=} &
    \psfig{file=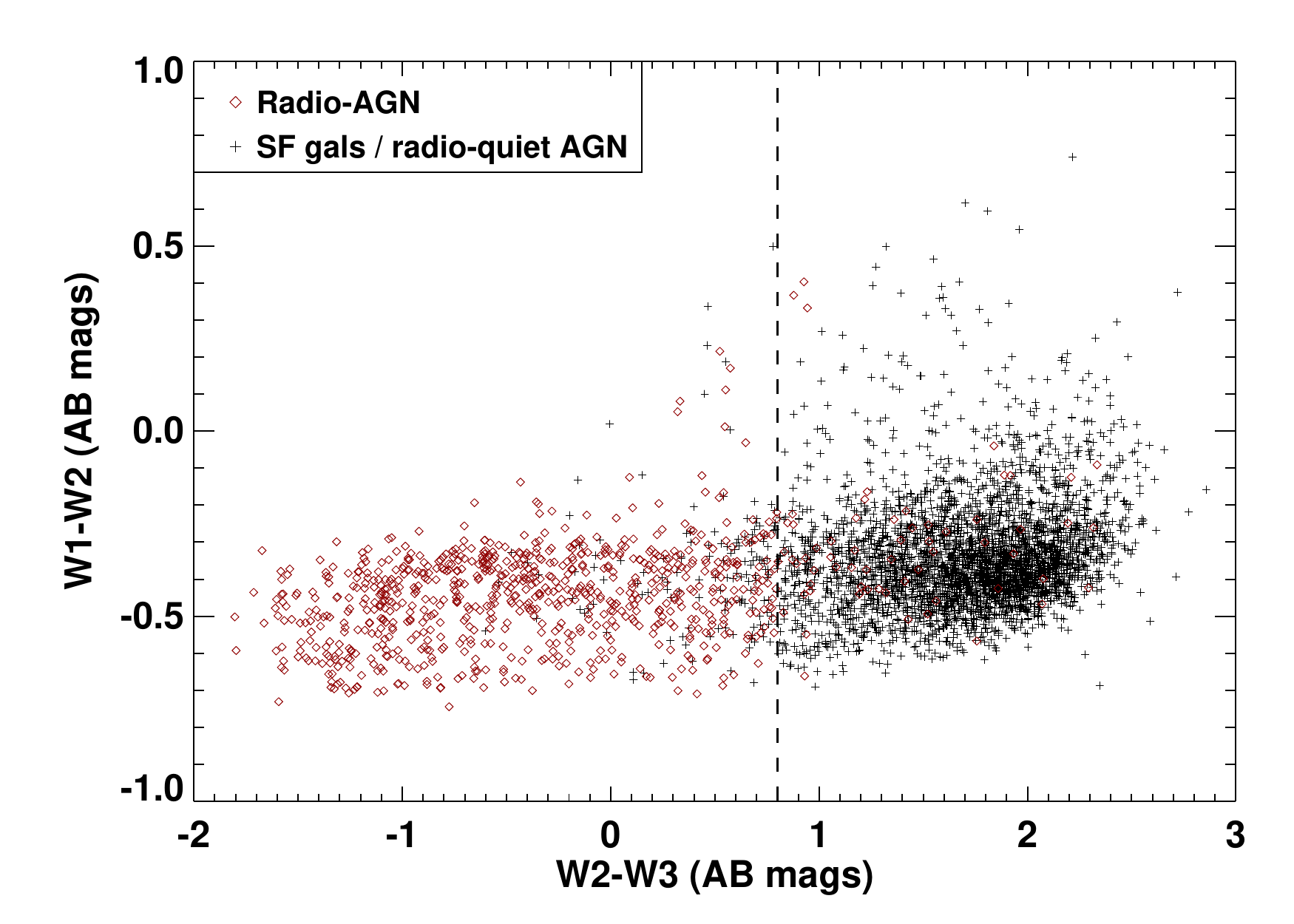,width=8.8cm,clip=} \\
  \end{tabular}
  \caption{Location of the LoTSS sources on the four diagnostic plots used to 
separate the radio AGN from those galaxies where the radio emission is powered 
by SF. {\it Upper left:} The `$D_{4000}$ versus $L_{\rm 150\,MHz}$/$M_*$' 
method, developed by \citet{Best2005a}. Upper right: The [O{\sc iii}]/H$\beta$ 
versus [N{\sc ii}]/H$\alpha$ emission line ratio diagnostic diagram of 
\citet{Baldwin1981}. Lower left: The $L_{\mathrm{H}\alpha}$ versus $L_{\rm 
150\,MHz}$ relation. Lower right: This WISE W1--W2 versus W2--W3 colour-colour 
diagnostic. In all plots, the dotted lines indicate the division(s) used for 
that classification method (see Section~\ref{sec:sfagndiags} for further 
details); in the cases of the upper left and lower left plots, there are two 
division lines; sources between the two lines are deemed to be `intermediate' by 
that classification method. The different symbols reflect the final combined 
classification of each source: radio AGN are plotted as red diamonds and SFGs as 
black crosses. In the lower left plot, arrows indicate upper limits to the 
H$\alpha$ luminosity. }
  \label{fig:sfagn}
\end{figure*}

Figure~\ref{fig:sfagn} shows the four classification methods along with the 
classification lines adopted. Sources are colour-coded by their final adopted 
classification (see Section~\ref{sec:final_class}). The details of the 
classification in each of the four diagnostics is now discussed in turn.

The upper left panel of Figure~\ref{fig:sfagn} shows $D_{4000}$ {\it versus} 
$L_{\rm 150\,MHz}$/$M_*$, based on that of \citet{Best2005b}. The upper dotted 
line shows the revised division proposed by \citet{Kauffmann2008} and adopted by 
\citet{Best2012}; the radio luminosity is converted to 150\,MHz from 1.4\,GHz by 
assuming a spectral index of $\alpha = 0.7$ (for $S_\nu \propto \nu^{-\alpha}$; 
this spectral index is the canonical value from \citealt{Condon2002}, with a 
similar typical value found for LOFAR sources by \citealt{CalistroRivera2017}). 
However, applying these classifications to the H-ATLAS data indicated that this 
conservative cut left many genuine radio AGN within the star-forming region and 
led to significant numbers of misclassifications in the final combined 
classifications. In contrast, \citep[as noted by][]{Kauffmann2008} the original 
cut of \cite{Best2005b} led to some SFGs being misclassified as radio AGN and, 
especially at the lower radio luminosities probed by LoTSS, also led to 
misclassification of some radio AGN with large values of $D_{4000}$ as SFGs. 
Therefore, a second diagnostic line was introduced that tracked the original 
\citet{Best2005b} line until $D_{4000} = 1.7$ and then continued horizontally 
(lower dotted line on the upper left panel of Figure~\ref{fig:sfagn}); this form 
was chosen to maximise agreement with the more sophisticated \citet{Gurkan2018} 
classifications of the (much smaller) H-ATLAS sample. Galaxies above the upper 
dotted line were classified as radio AGN via this diagnostic, galaxies below the 
lower dotted line were classified as SFGs, and galaxies between the two lines 
were deemed intermediate. Only a handful of galaxies could not be classified on 
this diagnostic because of an absence of a reliable mass measurement. 
Table~\ref{tab:sfagn2} shows the number of objects classified into each category 
by each diagnostic. It also shows the resultant overall classification 
statistics of these galaxies (see Section~\ref{sec:final_class}) and can thus be 
used to judge the importance of each diagnostic in the overall classification.

\begin{table}
\centering
\caption{Numbers of sources classified in each category by each different 
classification method. `AGN' and `SF' are (radio) AGN and star-forming 
classifications, respectively. `Intermediate' are the intermediate 
classifications for the $D_{4000}$ versus $L_{\rm 150\,MHz}$/$M_*$ or 
$L_{\mathrm{H}\alpha}$ versus $L_{\rm rad}$ diagnotics; `unclassified' sources 
lack data for classification on that diagnostic. The bracketed number on the 
second row for each diagnostic shows the number of these sources that end up 
with an overall classification of `radio AGN' (with the rest classified as 
star-forming, which may include radio-quiet AGN). This can therefore be used to 
gauge the importance of each diagnostic in the overall classification. For 
example, the majority of objects classified as `AGN' by the BPT diagnostic are 
ultimately deemed to be radio-quiet AGN so this classification is not given much 
weight, although classification as `SF' by the same diagnostic is almost 
invariably secure.}
\label{tab:sfagn2}
\begin{tabular}{ccccc}
  \hline
  Diagnostic & \multicolumn{4}{c}{Number classified in that category} \\
  method     & \multicolumn{4}{c}{(number with overall class of radio AGN)}\\
  & AGN & SF & Intermediate & Unclassified \\
\hline
\hline
$D_{4000}$ vs.    & 703 & 8392 & 1510 & 10 \\
$L_{\rm 150\,MHz}$/$M_*$     & (703) & (97) & (1318) & (3) \\
\hline
BPT                   & 2865 & 5012 & 0 & 2738 \\
                      & (544) & (10) & (0) & (1567) \\
\hline
$L_{\mathrm{H}\alpha}$ vs. & 659 & 8622 & 1146 & 188 \\
$L_{\rm 150\,MHz}$       & (659) & (532) & (753) & (177) \\
\hline
WISE                  & 1759 & 8039 & 0 & 781 \\
W2-W3                 & (1426) & (105) & (0) & (590) \\
\hline
\end{tabular}
\end{table} 

The upper right panel of Figure~\ref{fig:sfagn} shows the [O{\sc iii}]/H$\beta$ 
versus [N{\sc ii}]/H$\alpha$ line diagnostic diagram that is widely used to 
separate AGN and SFGs owing to the relative strength of the four emission lines 
involved and the broad independence of the line ratios on dust extinction. The 
division proposed by \citet{Kauffmann2003} is adopted: AGN and SFGs are divided 
at log([O{\sc iii}]/H$\beta$) $=$ 1.3 + 0.61 / (log([N{\sc ii}]/H$\alpha$) $-$ 
0.05). Seventy-four per cent of the radio source sample are classifiable by this 
method, while the rest lack detections for at least one of the lines. This is a 
significantly higher fraction of classifiable objects than the $\sim 30$ per 
cent found by \citet{Best2012}, largely because the deeper radio survey contains 
a far higher fraction of (stronger lined) SFGs. As discussed above (and evident 
on Figure~\ref{fig:sfagn}) objects selected as AGN by this diagnostic may be 
significantly contaminated by radio-quiet AGN, and so for these objects this 
diagnostic diagram is not given much weight in the final classification (cf. 
Table~\ref{tab:sfagn2}~and~\ref{tab:sfagn}). However, the absence of AGN 
signatures provides a more useful diagnostic: essentially all $\sim 5000$ 
sources ($=47$ per cent of the sample) that fall in the SFG region of this 
diagnostic plot have an overall classification of SFG (see 
Table~\ref{tab:sfagn2}).

The $L_{\mathrm{H}\alpha}$ versus $L_{\rm 150\,MHz}$ diagnostic is shown in the 
lower left panel of Figure~\ref{fig:sfagn}. \citet{Best2012} used just a single 
conservative cut on this diagnostic. However, as with the $D_{\rm 4000}$ versus 
$L_{\rm 150\,MHz}$/$M_*$ diagnostic, comparison with the H-ATLAS data indicates 
that any single separation line leads to significant numbers of 
misclassifications. Therefore, again, two separation lines are adopted: $\log 
(L_{\mathrm{H}\alpha} / \mathrm{L_{\odot}}) = \log (L_{\rm 150\,MHz} / {\rm 
W\,Hz}^{-1}) - 16.9$ and $\log (L_{\mathrm{H}\alpha} / \mathrm{L_{\odot}}) = 
\log (L_{\rm 150\,MHz} / {\rm W\,Hz}^{-1}) - 16.1$. Galaxies with measured 
H$\alpha$ luminosity, or an upper limit on this, below the lower line are 
classified as radio AGN on this diagnostic. Those which have measured H$\alpha$ 
luminosities, or upper limits, between the two lines are deemed intermediate, 
and those with a measured H$\alpha$ luminosity above the upper line are 
classified as SFGs. We found 1.7 per cent of sources have upper limits on 
$L_{\mathrm{H}\alpha}$ above the upper line and are left unclassified.

Finally, the lower right panel of Figure~\ref{fig:sfagn} shows a plot of W1--W2 
versus W2--W3 mid-infrared WISE colours. The sources are classified on this plot 
according to a simple division at W2--W3 = 0.8 (AB). This value was again 
optimised based on comparison with the H-ATLAS sample and is very similar to the 
division adopted by \citet{Herpich2016}. We found 7 per cent of sources did not 
possess a W2--W3 colour and so were unclassified. This diagnostic is somewhat 
crude, and was largely only used where the other diagnostics produced 
intermediate or contradictory classifications.

\subsection{Combination of diagnostics and final classification}
\label{sec:final_class}

For each individual classification method, sources may be classified as a radio 
AGN, classified as having their radio emission associated with SF, or be 
unclassified. For two of the four diagnostics, an intermediate classification is 
also possible. This leads to 144 different possible combinations of 
classifications, where in some cases the classifications may disagree. 
\citet{Best2012} discussed how to combine these individual classifications in 
order to obtain a single overall classification. Their classifications are 
broadly adopted here, but need to be expanded to incorporate the new 
intermediate classes and the WISE diagnostics. In this work, the approach taken 
was to apply these diagnostics to the equivalent data for the H-ATLAS sample 
from \citet{Gurkan2018}, and compare against  the more sophisticated 
classifications available for that sample. Specifically, for each possible 
combination of the four diagnostics, the H-ATLAS sources with that combination 
were identified, and the majority classification of that sample (SF or 
radio AGN) was adopted as the overall classification for that combination. In 
most cases this outcome was very clear with the H-ATLAS sources having (almost) 
uniformly the same classification; only for a few of the (less common) 
combinations, where the different diagnostic methods suggest different 
classifications, was the H-ATLAS sample also more heterogeneous in its 
classifications, suggesting (as expected) more uncertainty in the  overall 
classification. Based on this comparison with H-ATLAS, the overall potential 
contamination from misclassification is estimated to be $\la 3$ per cent. 

Table~\ref{tab:sfagn} shows the final classifications adopted and the number of 
sources in each category; to save space, classifications with fewer than five 
sources are not listed separately, but the overall contribution of these is 
indicated at the bottom of the table. In total, 2121 sources are classified as 
radio AGN, and 8494 sources are classified as having their radio emission 
associated with SF\footnote{These AGN/SF classifications are used in 
\citet{Hardcastle2018} to help to establish criteria to define larger AGN and SF 
galaxy samples from the wider LoTSS sample without SDSS spectroscopy.}.

\begin{table}
\centering
\caption{Number of sources and overall classification for different combinations 
of the four classification methods (with at least five sources).  `AGN' are 
sources classified as (radio) AGN; `SF' means that the radio emission is 
identified to be primarily powered by star formation (but a radio-quiet AGN may 
also be present); `Int' are intermediate classifications for the $D_{4000}$ 
versus $L_{\rm 150\,MHz}$/$M_*$ or $L_{\mathrm{H}\alpha}$ versus $L_{\rm rad}$ 
diagnotics; `Unc' (unclassified) sources lack data for classification on that 
diagnostic.}
  \label{tab:sfagn}
\begin{tabular}{cccccc}
\hline
 $D_{4000}$ vs. & BPT & $L_{\mathrm{H}\alpha}$ vs. & WISE & No. of & 
Overall\\
 $L_{\rm 150\,MHz}$/$M_*$ & & $L_{\rm 150\,MHz}$ & W2--W3 & sources & 
class  \\
\hline
\hline
Unc  & SF   & SF   & SF   &    5 & SF   \\
SF   & Unc  & Unc  & SF   &   11 & SF  \\
SF   & Unc  & Unc  & AGN  &    5 & AGN  \\
SF   & Unc  & SF   & Unc  &    8 & SF  \\
SF   & Unc  & SF   & SF   &  892 & SF  \\
SF   & Unc  & SF   & AGN  &  102 & SF  \\
SF   & Unc  & AGN  & Unc  &    5 & AGN  \\
SF   & Unc  & AGN  & SF   &   36 & AGN  \\
SF   & Unc  & Int  & Unc  &    8 & SF  \\
SF   & Unc  & Int  & SF   &   83 & SF  \\
SF   & Unc  & Int  & AGN  &   21 & AGN  \\
SF   & SF   & SF   & Unc  &  100 & SF  \\
SF   & SF   & SF   & SF   & 4791 & SF  \\
SF   & SF   & SF   & AGN  &   25 & SF  \\
SF   & SF   & Int  & Unc  &    5 & SF  \\
SF   & SF   & Int  & SF   &   63 & SF  \\
SF   & AGN  & SF   & Unc  &   29 & SF  \\
SF   & AGN  & SF   & SF   & 1758 & SF  \\
SF   & AGN  & SF   & AGN  &  240 & SF  \\
SF   & AGN  & Int  & Unc  &   10 & AGN  \\
SF   & AGN  & Int  & SF   &  179 & SF  \\
SF   & AGN  & Int  & AGN  &   14 & AGN  \\
AGN  & Unc  & Unc  & Unc  &    8 & AGN  \\
AGN  & Unc  & AGN  & Unc  &  211 & AGN  \\
AGN  & Unc  & AGN  & AGN  &  251 & AGN  \\
AGN  & Unc  & Int  & Unc  &   65 & AGN  \\
AGN  & Unc  & Int  & AGN  &   70 & AGN  \\
AGN  & AGN  & AGN  & Unc  &    8 & AGN  \\
AGN  & AGN  & AGN  & SF   &    5 & AGN  \\
AGN  & AGN  & AGN  & AGN  &   43 & AGN  \\
AGN  & AGN  & Int  & Unc  &    5 & AGN  \\
AGN  & AGN  & Int  & SF   &    5 & AGN  \\
AGN  & AGN  & Int  & AGN  &   20 & AGN  \\
Int  & Unc  & Unc  & Unc  &   67 & AGN  \\
Int  & Unc  & Unc  & AGN  &   92 & AGN  \\
Int  & Unc  & SF   & Unc  &   38 & SF  \\
Int  & Unc  & SF   & SF   &   28 & SF  \\
Int  & Unc  & SF   & AGN  &  194 & AGN  \\
Int  & Unc  & AGN  & Unc  &   45 & AGN  \\
Int  & Unc  & AGN  & AGN  &   43 & AGN  \\
Int  & Unc  & Int  & Unc  &  131 & AGN  \\
Int  & Unc  & Int  & SF   &   49 & AGN  \\
Int  & Unc  & Int  & AGN  &  256 & AGN  \\
Int  & SF   & SF   & SF   &    7 & SF  \\
Int  & SF   & SF   & AGN  &    8 & AGN  \\
Int  & AGN  & SF   & Unc  &   17 & AGN  \\
Int  & AGN  & SF   & SF   &   64 & SF  \\
Int  & AGN  & SF   & AGN  &  312 & AGN  \\
Int  & AGN  & Int  & Unc  &   15 & AGN  \\
Int  & AGN  & Int  & SF   &   50 & SF  \\
Int  & AGN  & Int  & AGN  &   86 & AGN  \\
\hline
\multicolumn{4}{c}{Other combinations} & 24 & AGN \\
\multicolumn{4}{c}{Other combinations} &  8 &  SF \\
\hline
\multicolumn{4}{c}{Total} & 2121 & AGN \\
\multicolumn{4}{c}{Total} & 8494 &  SF \\
\hline
\end{tabular}
\end{table}

\section{Spectral indices of local radio AGN}
\label{sec:spectral_index}

It is well established that the low-luminosity radio AGN have typically much 
smaller physical sizes that their more luminous counterparts. \citet{Best2005a} 
found that the vast majority of radio sources in their SDSS-NVSS-FIRST 
cross-match are unresolved at the $\approx 5 \arcsec$ resolution of the FIRST 
survey \citep[see also][]{Baldi2010}. Higher resolution radio studies have 
confirmed these sources to be AGN, but have shown that they either remain 
unresolved down to sub-arcsecond (sub-kpc) resolution or present weak jets on 
scales of at most a few kiloparsec \citep{Baldi2015}. The core radio 
morphologies of these sources and their host galaxy properties largely resemble 
those of the more powerful extended radio sources, but they lack the extended 
emission; the explanation for this is still a matter of debate \citep[e.g.][and 
references therein]{Baldi2018}. For our radio-AGN sample, 28 per cent of the 
sources show multiple components and 15 per cent can be modelled by a single 
Gaussian but are resolved according to the criteria shown by 
\citet{Shimwell2018}. The remaining 57 per cent of the sources are unresolved 
single components, although this figure raises quickly with decreasing flux 
density.

A characteristic property of these radio AGN that can offer clues as to their 
nature is the radio spectral index: compact radio cores and hotspots display 
much flatter spectra than extended radio components. \citet{deGasperin2018} have 
measured the spectral index properties of radio sources from wide-area sky 
surveys and have found evidence that these flattened towards lower radio flux 
densities; similar flattening at lower flux densities has been seen in deeper 
studies of small sky areas \citep[e.g.][]{Prandoni2006} and is often ascribed to 
lower luminosity AGN. However, the spectral indices of these local very 
low-luminosity AGN have only been investigated for small samples 
\citep{deGasperin2011}, indicating a mix of flat and steep spectrum sources.

\begin{figure}
   \centering
\includegraphics[width=\linewidth]{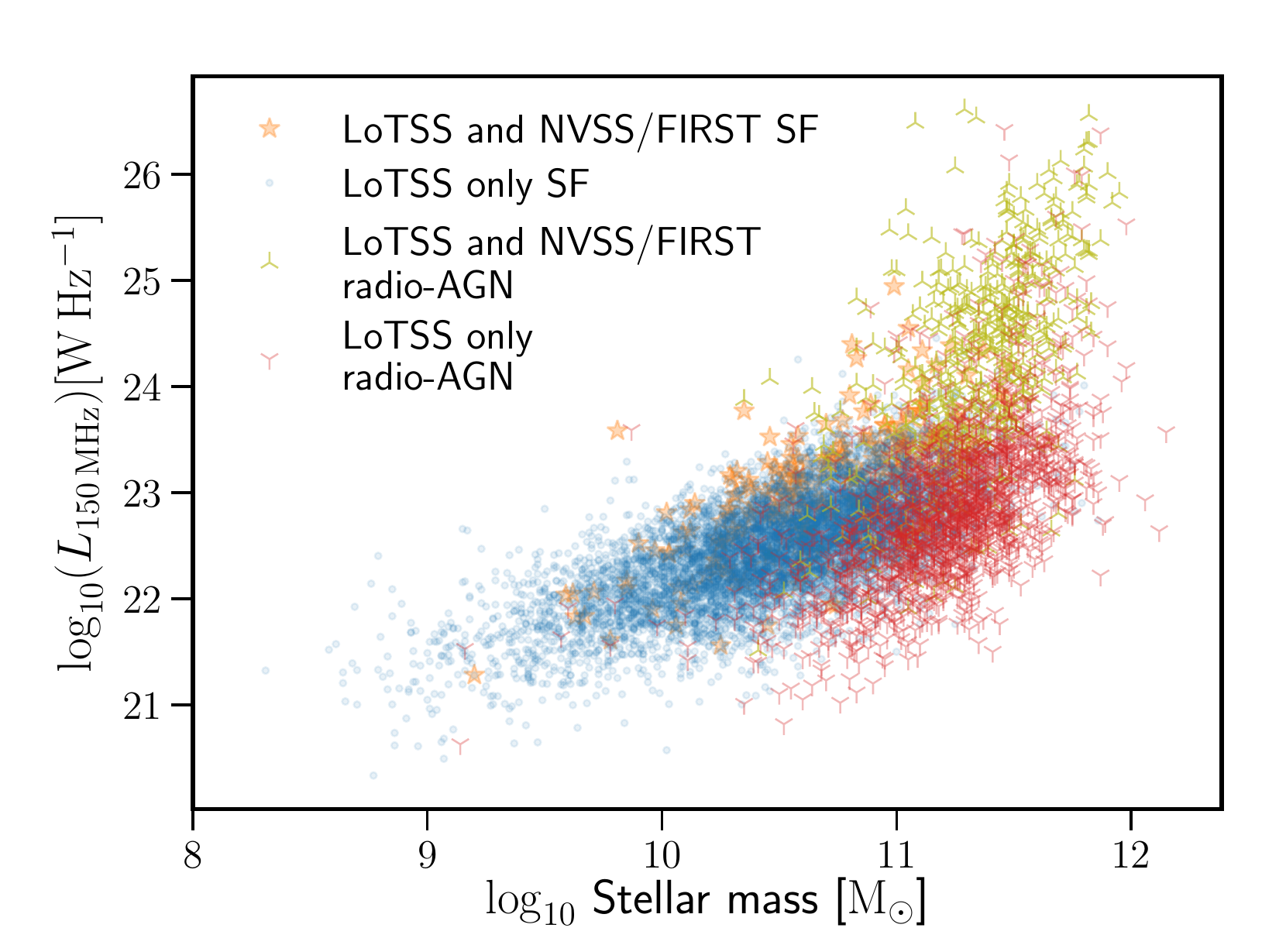}
   \caption{Distribution of LoTSS luminosities with respect to the stellar 
masses. The SF galaxies detected only in LoTSS are plotted with different 
symbols (circles) than those that are also detected in NVSS/FIRST (stars). Radio 
AGN are indicated with three-spoked asterisks in red for galaxies detected only 
in LoTSS and in olive for galaxies detected in both LoTSS and NVSS/FIRST.}
   \label{fig:mlum}
\end{figure}

Of the 2121 LoTSS-selected radio AGN, 496 of these sources are in common with 
the NVSS-FIRST sample of \citet{Best2012}, and therefore have available 1.4\,GHz 
flux densities. The cross-matching process of \citet[][see also 
\citealt{Best2005a}]{Best2012} uses an approach that combines NVSS and FIRST 
data, such that the matched sources benefit from the angular resolution of FIRST 
(which is similar to that of LoTSS) but also the sensitivity to extended 
structure offered by NVSS (which LoTSS also possesses due to the dense LOFAR 
core). This combination thus minimises any possible biases in spectral indices 
originating from a mismatch in resolution/sensitivity of different surveys. The 
relation between LoTSS luminosity and stellar mass, including which galaxies are 
also detected in NVSS/FIRST, is shown in Fig.~\ref{fig:mlum}. In this figure it 
is clear how LoTSS probes a population of galaxies that remained undetected in 
NVSS/FIRST.

\begin{figure}
   \centering
\includegraphics[width=\linewidth]{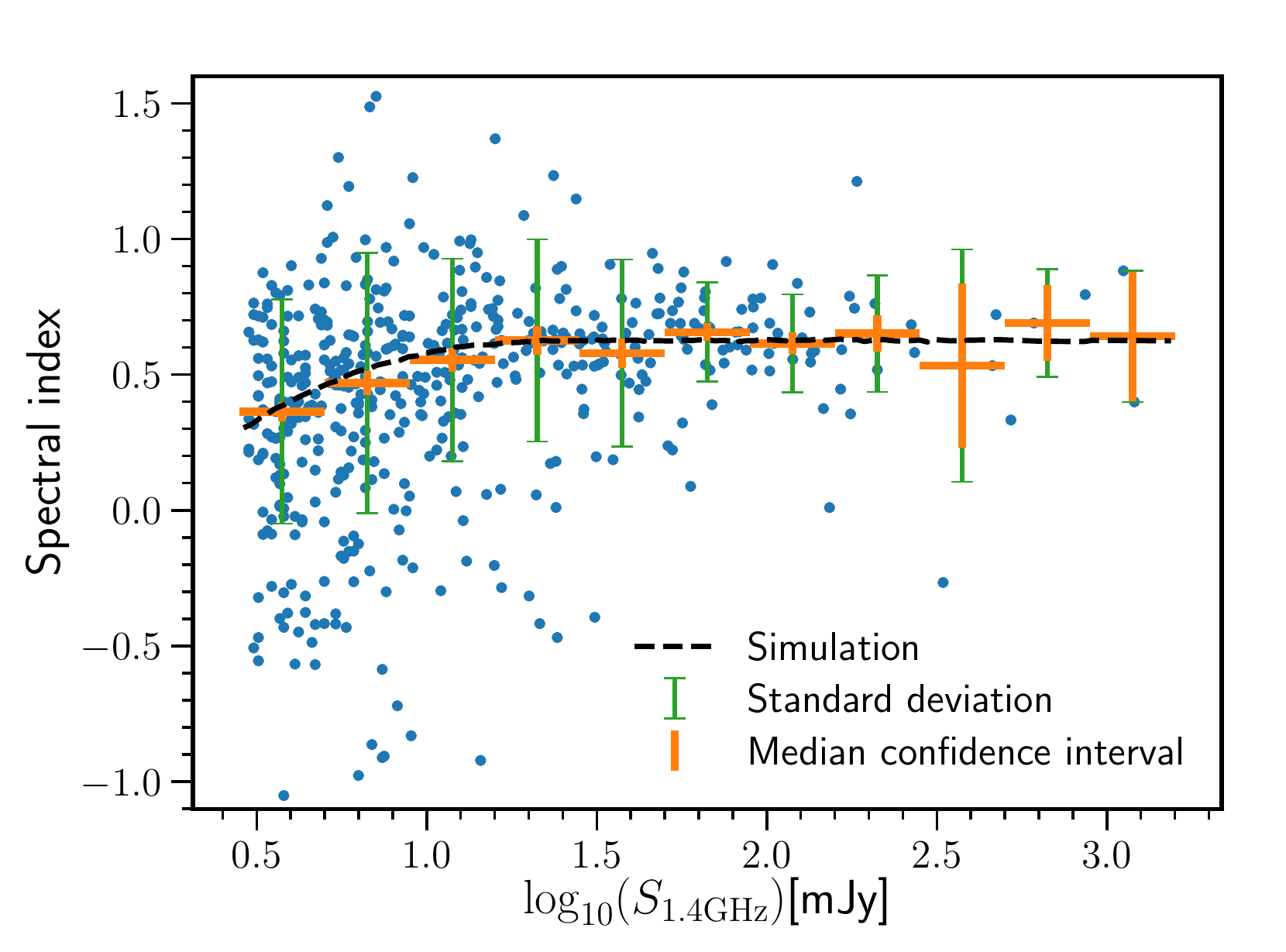}
   \caption{Spectral index distribution of the radio-AGN galaxies with both 
LoTSS and 1.4\,GHz measurements, as a function of 1.4\,GHz flux density. 
Individual sources are shown as blue dots and the medians of different bins in 
1.4\,GHz flux density are shown as orange crosses. The vertical lines correspond 
to the standard deviation on each bin. The median spectral index at flux 
densities $S_{\rm 1.4\,GHz} > 20$\,mJy is 0.63. The weak decrease in median 
spectral index at lower flux densities may be driven by selection biases in the 
1.4\,GHz sample, as indicated by the black dashed line that simulates the effect 
of this (see text for details).}
   \label{fig:alpha}
\end{figure}

The distribution of 150\,MHz to 1.4\,GHz spectral indices for these radio AGN is 
shown in Fig.~\ref{fig:alpha}, as a function of 1.4\,GHz flux density; the LoTSS 
data are sufficiently deeper than the NVSS/FIRST data that the LoTSS limit 
provides essentially no constraint on this plot. At high flux densities ($S_{\rm 
1.4\,GHz} > 20$\,mJy) the median spectral index is $0.63$, which is very much in 
line with the canonical value of 0.7 \citep{Condon2002}. The scatter of the 
spectral indices at the lower flux densities are mainly caused by the 
uncertainty in the measurements but a possible general trend towards flat or 
even inverted spectrum sources was also explored. 

The median spectral index decreases towards 0.4 at the lowest flux densities, 
and the faintest radio sources display a wide range of spectral indices, 
including some flat spectrum sources. If correct this would be consistent with 
the suggestion of \citet{Whittam2017}, using higher frequency source counts, of 
an increasing core fraction towards lower flux densities. However, it must be 
noted that at the lowest flux densities the distribution may be skewed by 
selection biases in the FIRST-NVSS sample as the 3\,mJy flux density limit of 
that sample (set by the NVSS limits) is approached, boosting the number of 
apparently flatter spectrum sources. These selection biases were investigated 
using a Monte Carlo simulation. The distribution of spectral indices for sources 
with $S_{\rm 1.4\,GHz} > 20$\,mJy was fitted using a Gaussian. Then, for each 
source with a flux density measured at 150\,MHz, 10000 random spectral indices 
were drawn from the Gaussian and the corresponding $S_{\rm 1.4\,GHz}$ was 
calculated. If this was below the NVSS detection limit (3\,mJy) then that 
iteration of that source was discarded (as it would not have been within the 
subsample of sources with measured spectral indices); otherwise it was retained. 
The retained values were then used to compute the mean spectral index as a 
function of 1.4\,GHz flux density. The results of the simulation are plotted in 
Fig.~\ref{fig:alpha}. It is clear that the trend of the spectral index to 
decrease at the lower flux densities is consistent with being driven entirely by 
the biases arising from the combination of statistical flux errors and selection 
limits. That would indicate that the bulk of the compact radio AGN are not 
simply flat spectrum radio cores with an absence of extended emission, but 
rather that they are broadly scaled-down versions of the more luminous extended 
sources \citep[see also discussion in][]{Baldi2018}. This conclusion fits in 
with the view of the recurrent nature of radio-AGN activity, discussed in 
Section~\ref{sec:accretion_rates}.

The typical spectral index found in Figure~\ref{fig:alpha} is also needed to 
provide a reliable conversion factor for comparison of the LoTSS results with 
previous measurements at higher frequencies.  For the remainder of the paper 
(and for consistency with the value adopted for the AGN/SF separation) the 
canonical spectral index for radio AGN of 0.7 is adopted, which is close to the 
observed median. This spectral index corresponds to the following conversion in 
luminosities: $\log_{10}(L_{\mathrm{1.4\,GHz}}) \approx \log_{10}(L_{\rm 
150\,MHz})-0.68$.


\section{Local 150\,MHz luminosity function}
\label{sec:lum_func}

The local 150\,MHz radio luminosity functions for both SFGs and radio AGN were 
determined using the standard technique, $\rho = \sum_i 1/V_i$ 
\citep{Schmidt1968,Condon1989}, where $V_i$ is the volume within which source 
$i$ could be detected. The calculation of $V_i$ for each source requires careful 
consideration of both the radio and optical redshift limits, in particular 
taking into account the variable flux limit of the LoTSS survey as a function of 
sky location \citep{Shimwell2018}.

For each source, $V_i$ was calculated as follows. First, the redshift range $z < 
0.3$ was divided into narrow redshift slices (in practice, 3000 slices of width 
$\Delta z = 0.0001$ were used). The radio luminosity of the source in question 
was then used to calculate the flux density that the source would have at the 
mid-point of each redshift slice, assuming a spectral index of 0.7. The LoTSS 
rms maps were used to calculate the sky area over which a source of that flux 
density could be detected above the 5$\sigma$ limit \citep[see][]{Shimwell2018}, 
and hence the volume available to that source within the redshift slice. 
Finally, the optical magnitude limits of the SDSS main galaxy selection ($14.5 < 
r < 17.77$) were used to determine upper and lower redshift limits at which the 
source could be included in the SDSS sample (if these magnitude limits lay 
outside the sample selection limits of $z_{\rm min}=0.01$ and $z_{\rm max} = 
0.3$ then the latter were used instead). The available volume in all redshift 
slices between those two limits was then summed. The radio luminosity functions 
were then derived by summing $1/V_i$ across all sources, and Poisson statistics 
were used to estimate the uncertainties.  It should be emphasised that at some 
luminosities the formal statistical uncertainties are small and are likely to be 
underestimates, and systematic errors such as AGN/SF classification dominated 
the error budget.  No correction is made for incompleteness in the LoTSS survey, 
but as shown by \citet{Shimwell2018} this is small above the 5$\sigma$ threshold 
(at least for point sources, which the low-luminosity radio AGN are likely to 
be). It may possibly contribute to the slight downturn in the faintest 
luminosity bin.

\begin{figure}   
   \centering
\includegraphics[width=\linewidth]{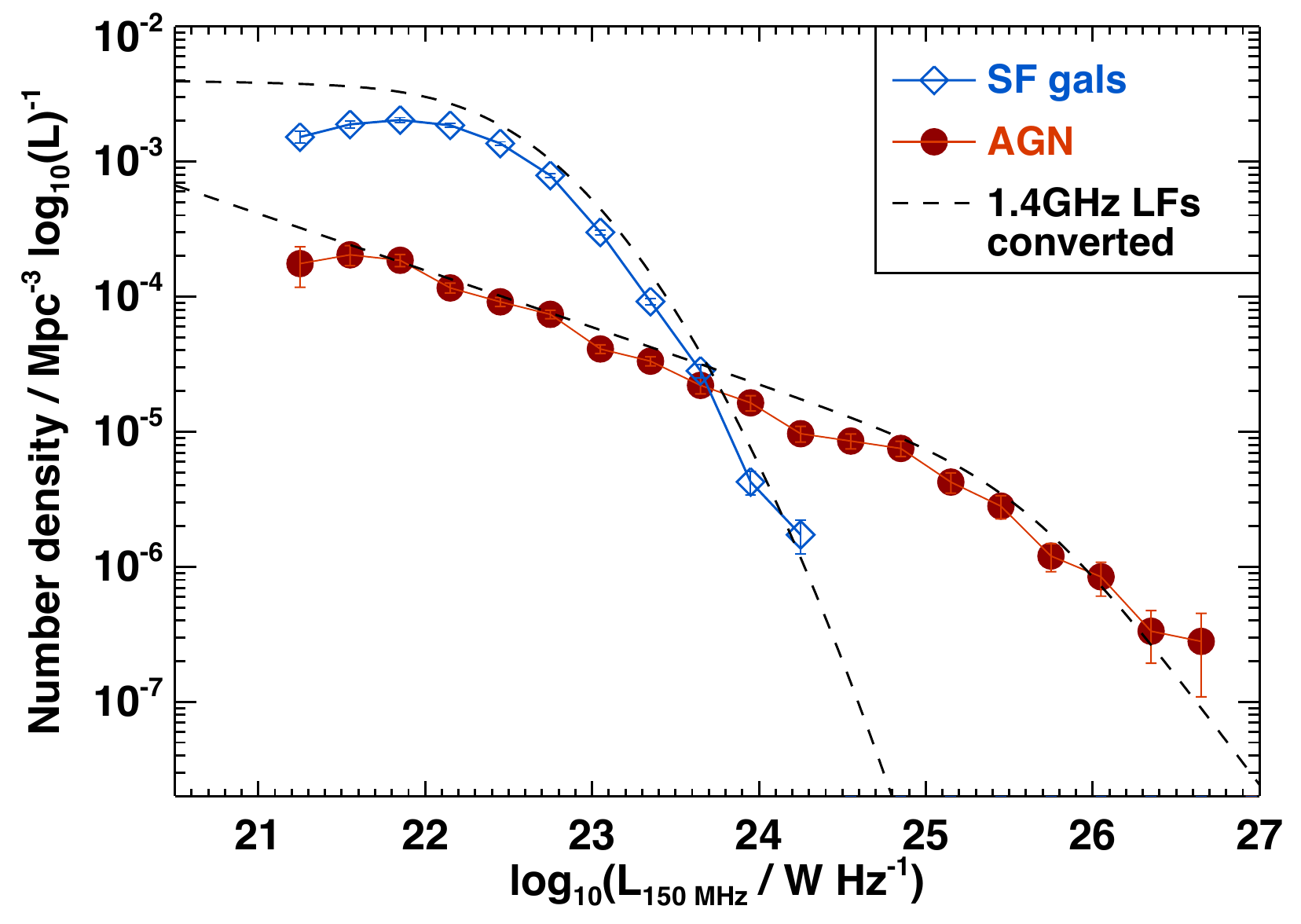}
   \caption{ Local 150\,MHz luminosity functions of AGN (red circles) and SF 
galaxies (blue diamonds) separately. Also shown for comparison (black dashed 
lines) are the local 1.4\,GHz luminosity function for radio AGN \citep[using the 
parameterisation of][]{Heckman2014} and SF galaxies \citep[using the 
parameterisation of][]{Mauch2007}, converted to 150\,MHz by scaling the break 
frequency with a spectral index of 0.7.}
   \label{fig:lumfunc}
\end{figure}

The derived radio luminosity functions of AGN and SF galaxies are provided 
separately in Table~\ref{tab:lumfunc} and shown in Figure~\ref{fig:lumfunc}.  
For comparison, the local 1.4\,GHz luminosity functions of radio AGN and SF 
galaxies are overlaid, converted to 150\,MHz by adjusting the break luminosity 
using the spectral index of 0.7. The plotted line for the radio AGN is the 
parameterisation of \citet{Heckman2014}, which averages over a wide selection of 
previous 1.4 GHz radio luminosity function determinations 
\citep{Machalski2000,Sadler2002,Best2005a,Mauch2007,Best2012}, while that of SF 
galaxies uses the parameterisation of \citet{Mauch2007}. As can be seen, the 
agreement between the 1.4\,GHz and 150\,MHz radio luminosity functions is good 
\citep[as is that with earlier 150\,MHz luminosity functions in the H-ATLAS 
field by][]{Hardcastle2016}. The slight offset in the SF galaxies can either be 
explained by the use of a different spectral index (the agreement above the 
break is almost perfect if $\alpha=0.6$ is used instead), or be caused by 
slightly different redshift distributions of the two samples, given the strong 
cosmological evolution of this population. These results give confidence in the 
robustness of the separation of SF galaxies from AGN, which is important for the 
analysis in the subsequent sections of the paper.

The luminosity function of radio AGN continues to increase with the same 
power-law slope down to the lowest flux densities probed by LoTSS. As argued by 
\citet{Mauch2007}, this cannot continue indefinitely without the integrated 
space density of radio AGN exceeding the space density of massive galaxies which 
are believed to host these objects. \citet{Mauch2007} calculated a limit of 
$L_{\rm 1.4\,GHz} \approx 10^{19.5}$\,W\,Hz$^{-1}$ below which the radio 
luminosity function must turn down if their hosts are all brighter than $L^*$ 
ellipticals, where $L^*$ is the break of the optical luminosity function, and 
\citet{Cattaneo2009b} derived a similar value assuming that the host galaxies 
require a black hole more massive than $10^6\,\mathrm{M_{\odot}}$. This limit 
corresponds to around $10^{20.2}$\,W\,Hz$^{-1}$ at 150\,MHz, less than an order 
of magnitude below the faintest luminosities probed by LoTSS. This issue is 
investigated in more detail in the following sections, by breaking down the 
radio-AGN prevalence by both (stellar or black hole) mass and radio luminosity 
separately.

\begin{table}
\caption{Local 150\,MHz luminosity functions of AGN and SF galaxies separately. 
The columns show the range of 150\,MHz radio luminosities considered in each 
bin, the total number of radio AGN in that bin, the derived space density of 
these (in units of number per log$_{10} L$ per Mpc$^3$), the total number of 
SFGs and their space density. Uncertainties are statistical Poissonian 
uncertainties only.}
  \label{tab:lumfunc}
\begin{tabular}{crcrc}
\hline
&
\multicolumn{2}{c}{Radio AGN} &
\multicolumn{2}{c}{Star-forming galaxies} \\
${\rm log}_{10} L_{\rm 150\,MHz}$ &
N & ${\rm log}_{10} \rho$ &
N & ${\rm log}_{10} \rho$ \\
{\footnotesize(W Hz$^{-1}$)} & 
& {\footnotesize (log$_{10}L)^{-1}$Mpc$^{-3}$} &
& {\footnotesize (log$_{10}L)^{-1}$Mpc$^{-3}$} \\
\hline
\hline
 21.10--21.40 &  12 & $-3.75^{+0.12}_{-0.17}$ &  120 & $-2.81^{+0.04}_{-0.04}$ 
\\
 21.40--21.70 &  40 & $-3.69^{+0.06}_{-0.07}$ &  353 & $-2.72^{+0.02}_{-0.02}$ 
\\
 21.70--22.00 & 105 & $-3.73^{+0.04}_{-0.04}$ &  811 & $-2.69^{+0.01}_{-0.01}$ 
\\
 22.00--22.30 & 171 & $-3.93^{+0.03}_{-0.03}$ & 1459 & $-2.73^{+0.01}_{-0.01}$ 
\\
 22.30--22.60 & 268 & $-4.03^{+0.03}_{-0.03}$ & 1935 & $-2.86^{+0.01}_{-0.01}$ 
\\
 22.60--22.90 & 332 & $-4.13^{+0.03}_{-0.03}$ & 1896 & $-3.10^{+0.01}_{-0.01}$ 
\\
 22.90--23.20 & 276 & $-4.38^{+0.03}_{-0.03}$ & 1114 & $-3.52^{+0.01}_{-0.01}$ 
\\
 23.20--23.50 & 251 & $-4.47^{+0.03}_{-0.03}$ &  487 & $-4.03^{+0.02}_{-0.02}$ 
\\
 23.50--23.80 & 151 & $-4.65^{+0.05}_{-0.06}$ &  165 & $-4.54^{+0.04}_{-0.05}$ 
\\
 23.80--24.10 & 118 & $-4.78^{+0.05}_{-0.06}$ &   38 & $-5.37^{+0.07}_{-0.09}$ 
\\
 24.10--24.40 &  81 & $-5.01^{+0.05}_{-0.06}$ &   14 & $-5.76^{+0.10}_{-0.14}$ 
\\
 24.40--24.70 &  83 & $-5.06^{+0.05}_{-0.05}$ &     &                     \\
 24.70--25.00 &  75 & $-5.12^{+0.05}_{-0.06}$ &     &                     \\
 25.00--25.30 &  48 & $-5.37^{+0.06}_{-0.08}$ &     &                     \\
 25.30--25.60 &  34 & $-5.55^{+0.07}_{-0.09}$ &     &                     \\
 25.60--25.90 &  19 & $-5.92^{+0.09}_{-0.11}$ &     &                     \\
 25.90--26.20 &  14 & $-6.07^{+0.10}_{-0.14}$ &     &                     \\
 26.20--26.50 &   6 & $-6.47^{+0.15}_{-0.23}$ &     &                     \\
 26.50--26.80 &   3 & $-6.55^{+0.20}_{-0.41}$ &     &                     \\
 \hline
\end{tabular}
\end{table}


\section{Fraction of radio AGN}
\label{sec:fractions}

\begin{figure*}   
   \centering
\raisebox{0.2cm}{\includegraphics[width=0.43\linewidth]
{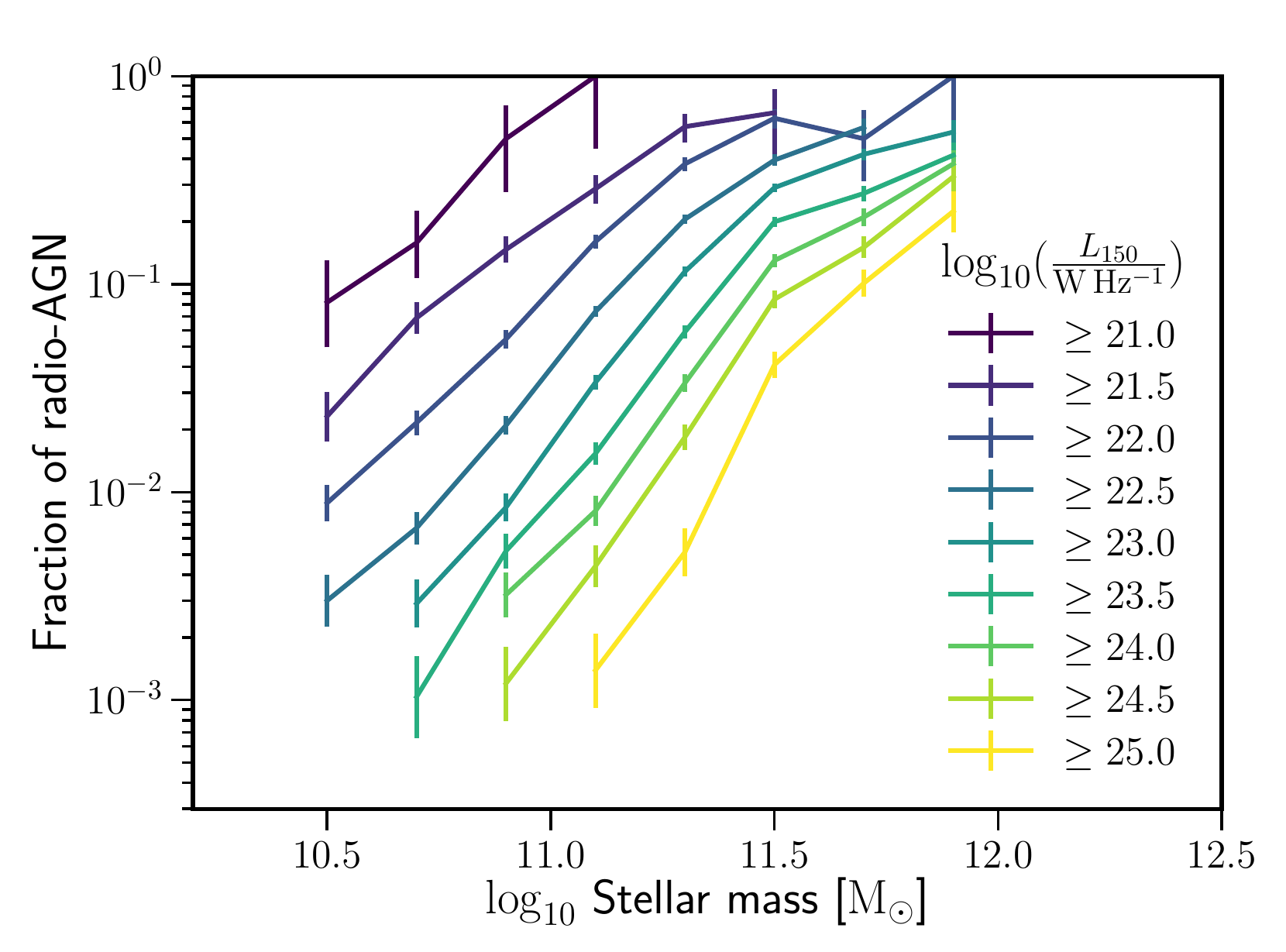}}
\includegraphics[width=0.55\linewidth]
{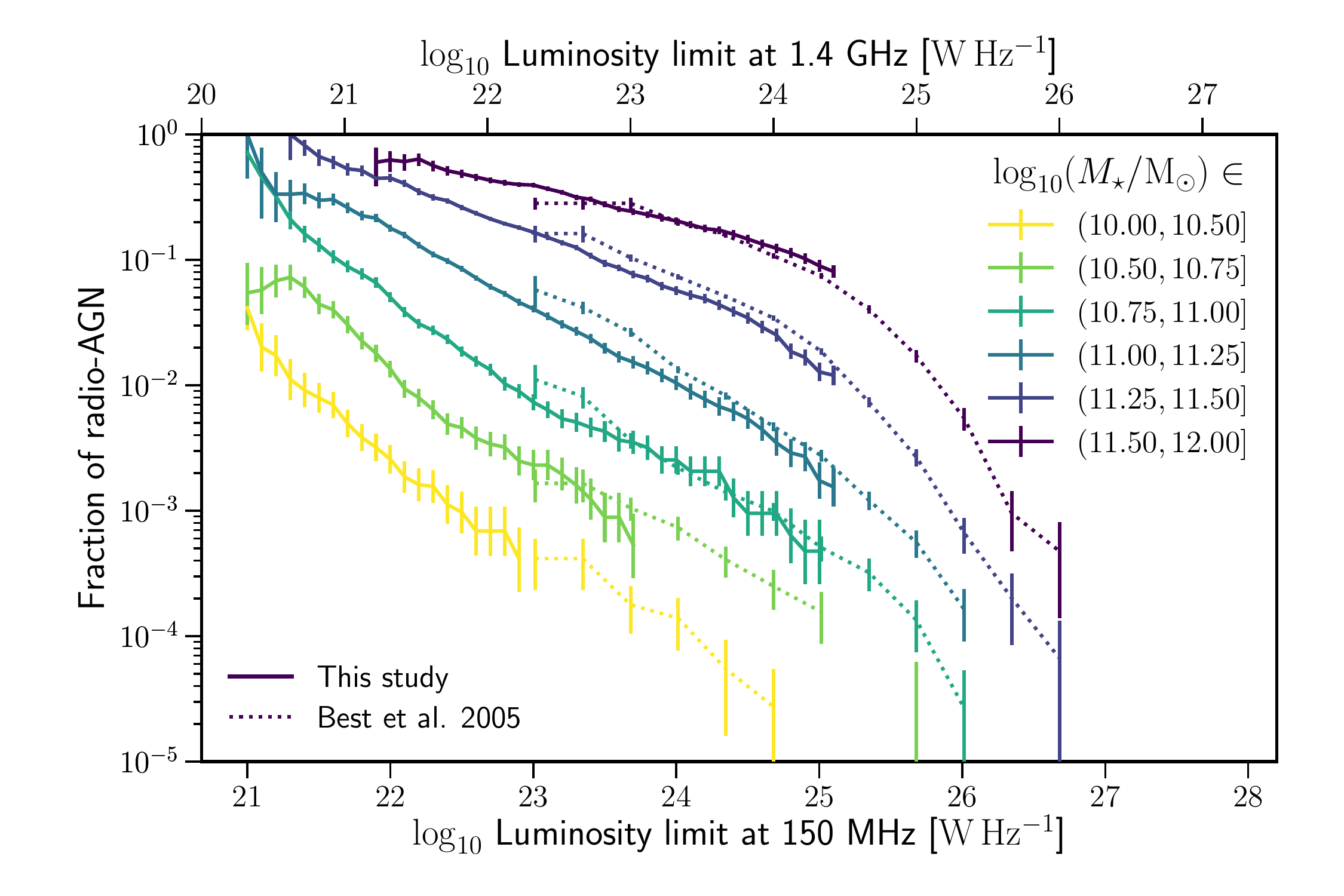}
   \caption{Left panel: Fraction of galaxies that host a radio AGN above a given 
luminosity limit, as a function of  stellar mass, for multiple radio luminosity 
limits (shown in different colours). Right panel: The fraction of galaxies 
hosting radio AGN brighter than a given radio luminosity, separated by their 
stellar mass. The various stellar mass ranges are shown in different colours. 
The solid lines represent the results of this (150\,MHz) study and dotted lines 
show the results from \citet{Best2005b} at 1.4\,GHz (converted assuming a 
spectral index of 0.7). The error bars correspond to a confidence interval 
equivalent to one sigma. The fractions of radio AGN are shown only up to 
$L_{\mathrm{150\,MHz}} \approx 10^{25}\,\mathrm{W\,Hz^{-1}}$ because the number 
of galaxies above this limit is relatively low and render very large error bars. 
The highest mass bin would reach a value compatible with 100 per cent within the 
error at $L_{\mathrm{150 MHz}} \leq 10^{21.7}\, \mathrm{W\,Hz^{-1}}$ but those 
points are not shown owing to the large error bars.}
   \label{fig:fractions_stellar}
\end{figure*}

It is well-established that there is a strong increase in the prevalence of 
radio-AGN activity with the mass of the galaxy \citep[e.g.][]{Best2005b, 
Mauch2007, Sabater2013}. The left panel of Fig.~\ref{fig:fractions_stellar} 
shows the fraction of the SDSS galaxies that host a radio AGN above a given 
luminosity limit, as a function of the stellar mass of the galaxy; this is shown 
for different cut-offs on the radio luminosity, from $L_{\mathrm{150\,MHz}} \geq 
10^{21}\,\mathrm{W\,Hz^{-1}}$ to $L_{\mathrm{150\,MHz}} \geq 
10^{25}\,\mathrm{W\,Hz^{-1}}$ in increments of 0.5 dex. For each limit, the 
fraction is computed by considering only the galaxies that could be detected 
above the given luminosity limit, with a flux density of five times the rms 
noise level. The error bars were computed using the Agresti-Coull binomial 
proportion confidence interval \citep{Agresti1998} with the confidence interval 
covering 68 per cent of the probability (roughly equivalent to 1 sigma in a 
normal distribution).  The results show the expected increase in the prevalence 
of radio AGN with stellar mass, as seen in previous studies at high 
luminosities, but are able to extend this down to lower radio luminosities. 
Remarkably, for limits of $L_{\mathrm{150\,MHz}} \geq 
10^{21}\,\mathrm{W\,Hz^{-1}}$, the fraction of galaxies hosting radio AGN 
reaches 100 per cent for stellar masses above $10^{11}$\,M$_{\odot}$. At limits 
below or equal to $L_{\mathrm{150\,MHz}} \geq 10^{22}\,\mathrm{W\,Hz^{-1}}$, the 
100 per cent fraction is reached for masses above $10^{12}$\,M$_{\odot}$. This 
is consistent with the results of \citet{Brown2011}, who studied a much smaller 
sample of nearby very massive galaxies and also found near ubiquity of radio 
emission from either AGN or SF activity.

The right panel of Fig.~\ref{fig:fractions_stellar} shows the same results in an 
alternative way. The fraction of galaxies hosting radio AGN is presented with 
respect to the luminosity limit for multiple strata of stellar mass. For 
comparison, this figure also shows the results of \citet{Best2005b} at higher 
luminosities at 1.4\,GHz, converted using a spectral index of 0.7. The results 
agree in the overlapping luminosities, but the current analysis extends almost 
two orders of magnitude lower in terms of radio luminosity. The new data discard 
the flattening towards lower luminosity limits that was hinted at in 
\citet{Best2005b} and \citet{Mauch2007}, instead indicating that the prevalence 
reaches 100 per cent by the lowest luminosities, at least for the higher mass 
ranges.

\begin{figure*}   
   \centering
\raisebox{0.2cm}{\includegraphics[width=0.43\linewidth]
{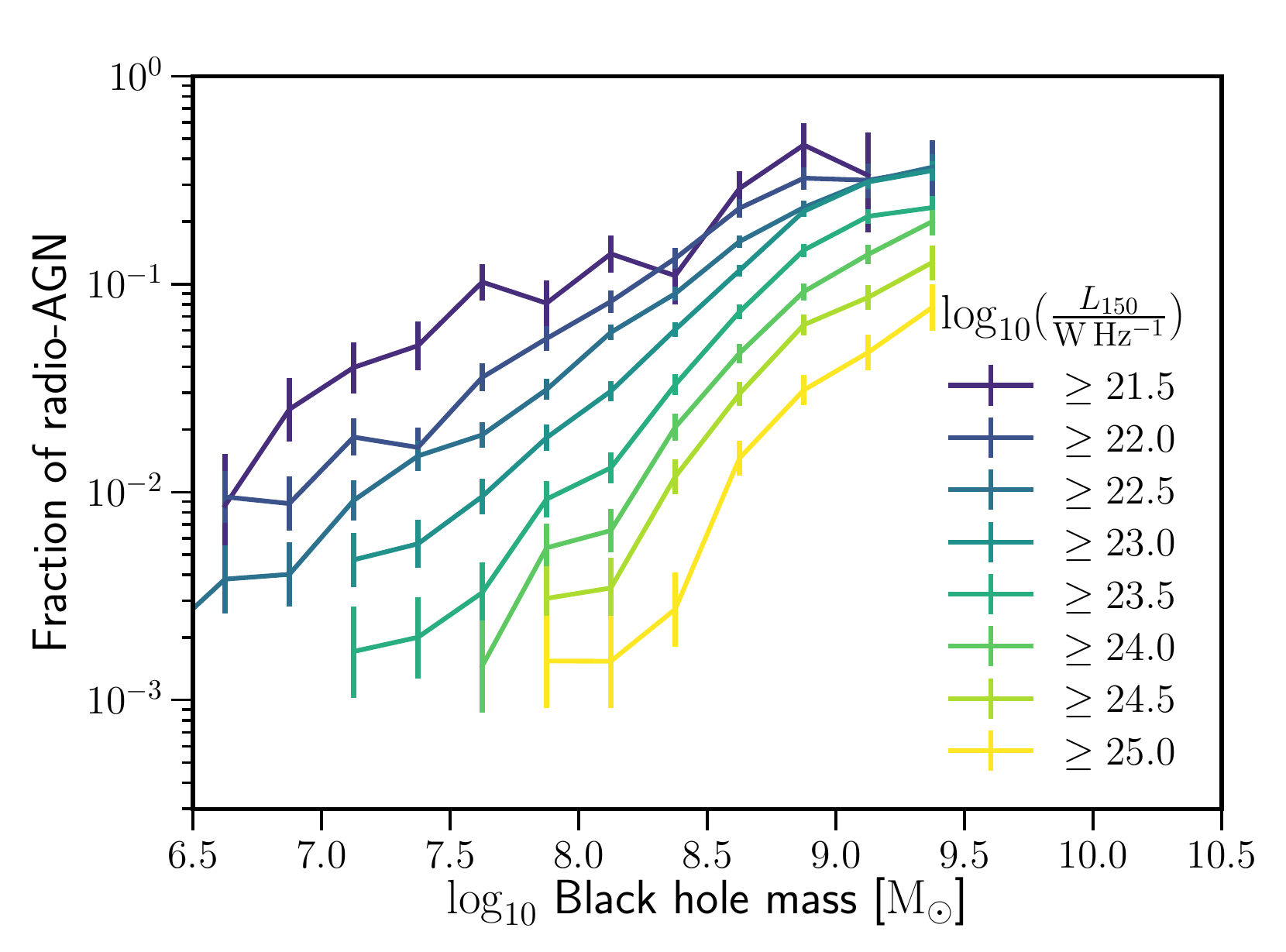}}
\includegraphics[width=0.55\linewidth]
{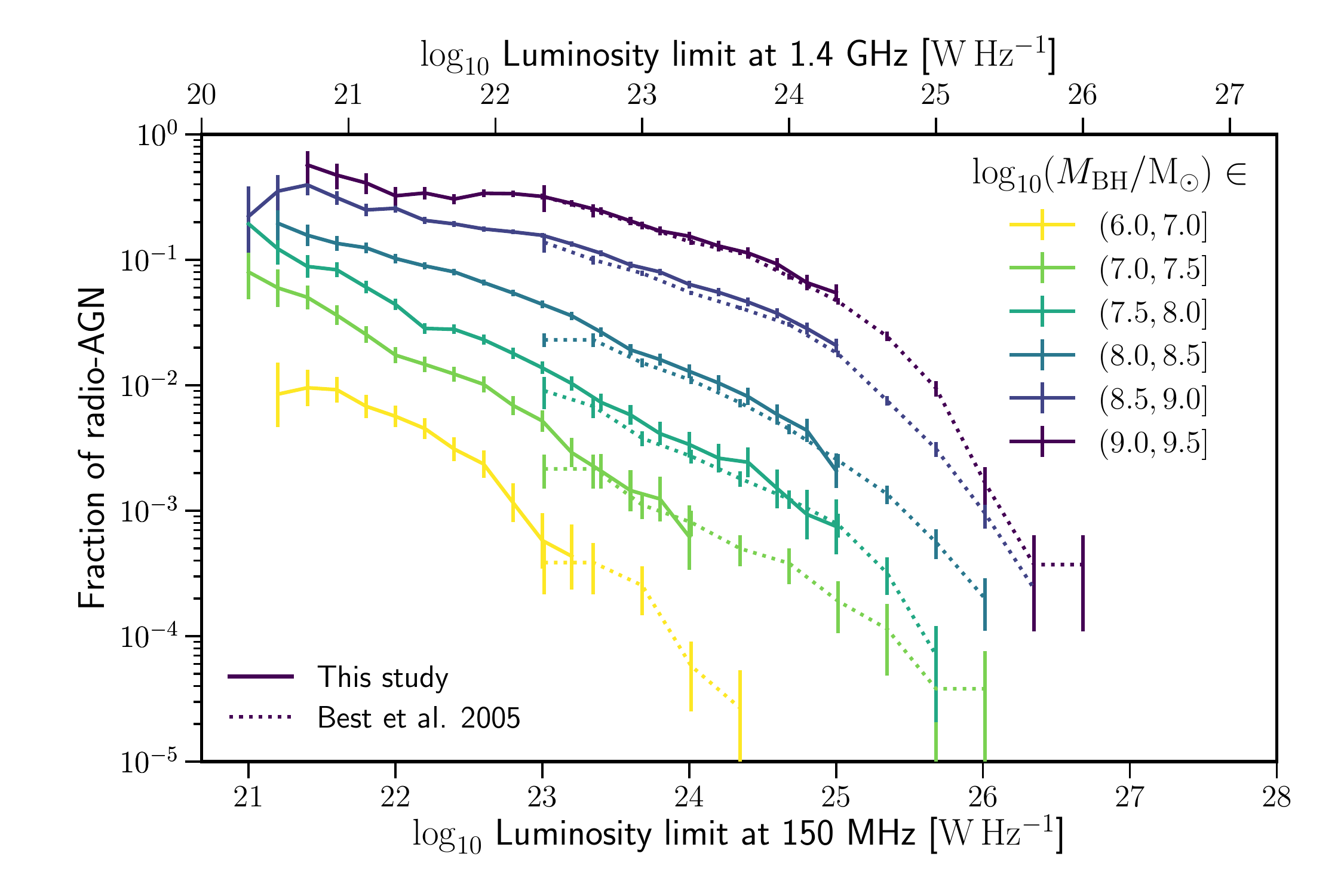}
   \caption{Left panel: Fraction of galaxies that host a radio AGN above a given 
radio luminosity, as a function of the black hole mass, for various radio 
luminosity limits (shown in different colours). Right panel: The fraction of 
galaxies hosting radio AGN brighter than a given radio luminosity, separated by 
their black hole mass. Colours, lines, and errors bars as in 
Figure~\ref{fig:fractions_stellar}. In the left panel, the line corresponding to 
$L_{\mathrm{150\,MHz}} \leq 10^{21}\,\mathrm{W\,Hz^{-1}}$ is omitted from the 
plot owing to its large error bars.}
   \label{fig:fractions_bh}
\end{figure*}

Fig.~\ref{fig:fractions_bh} considers the prevalence of radio-AGN activity in 
relation to the mass of the black hole; it is similar to 
Fig.~\ref{fig:fractions_stellar} but for black hole masses instead of stellar 
masses. In the left panel of Fig.~\ref{fig:fractions_bh} an increase in the 
fraction of radio AGN is seen with respect to the black hole mass. As with 
previous 1.4\,GHz studies, this has a shallower slope than found for stellar 
mass. Furthermore, in this case the slope appears to flatten at the highest 
black hole masses for each luminosity bin and the maximum fraction reaches at 
most 30 to 50 per cent. The right panel of Fig.~\ref{fig:fractions_bh} compares 
the results against the 1.4\,GHz data. For this comparison, the data of 
\citet{Best2005b} were re-analysed using new black hole masses derived using the 
velocity dispersion to black hole mass relation adopted in this paper (see 
Sect.~\ref{sec:sample_data}; this is different from the relation that 
\citeauthor{Best2005b} originally used). Once again, the agreement is excellent, 
and in this case the flattening persists and the fraction never reaches 100 per 
cent even for the highest black hole masses.

\begin{figure*}   
   \centering
\includegraphics[width=0.45\linewidth]
{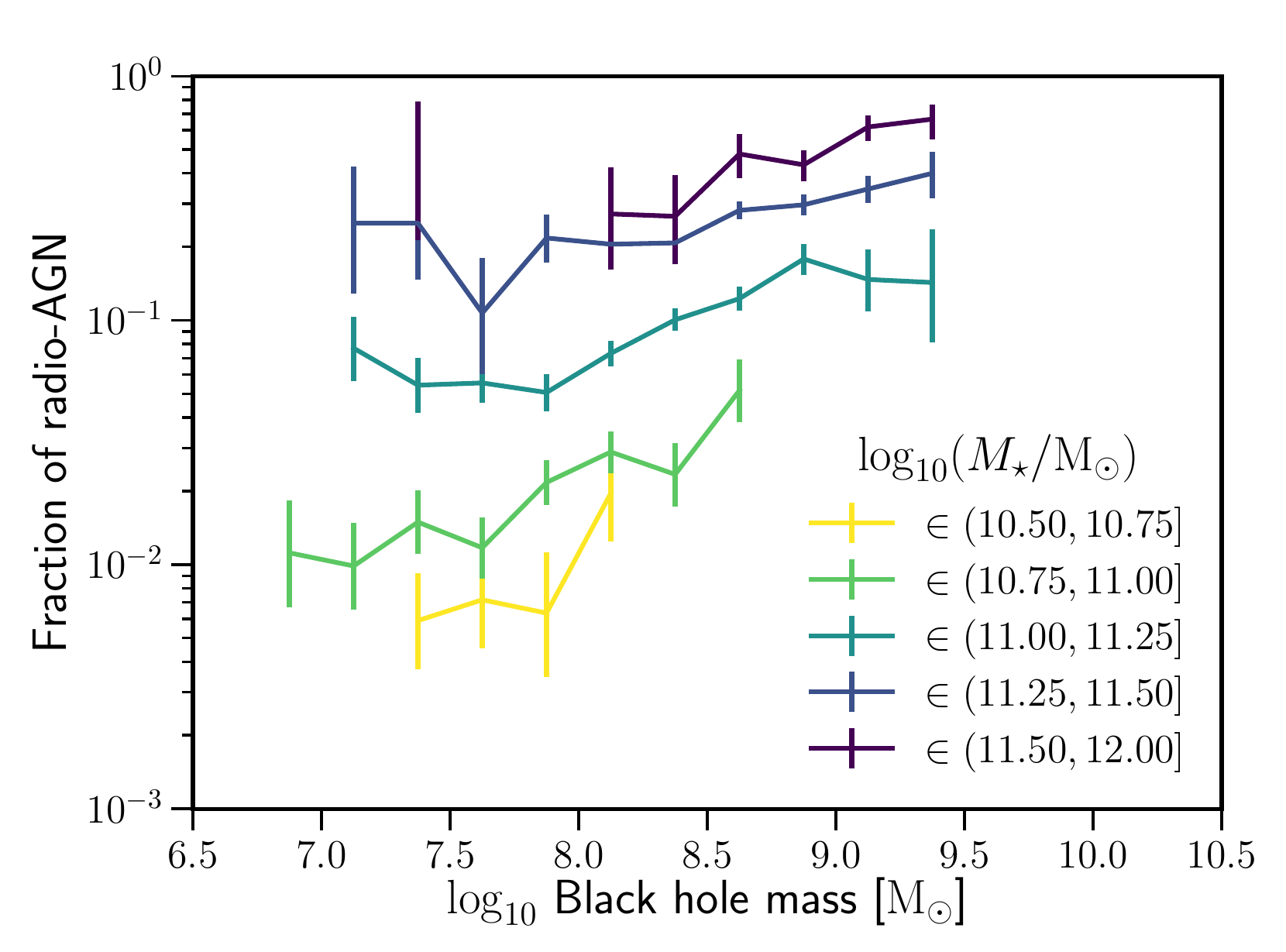}
\includegraphics[width=0.45\linewidth]
{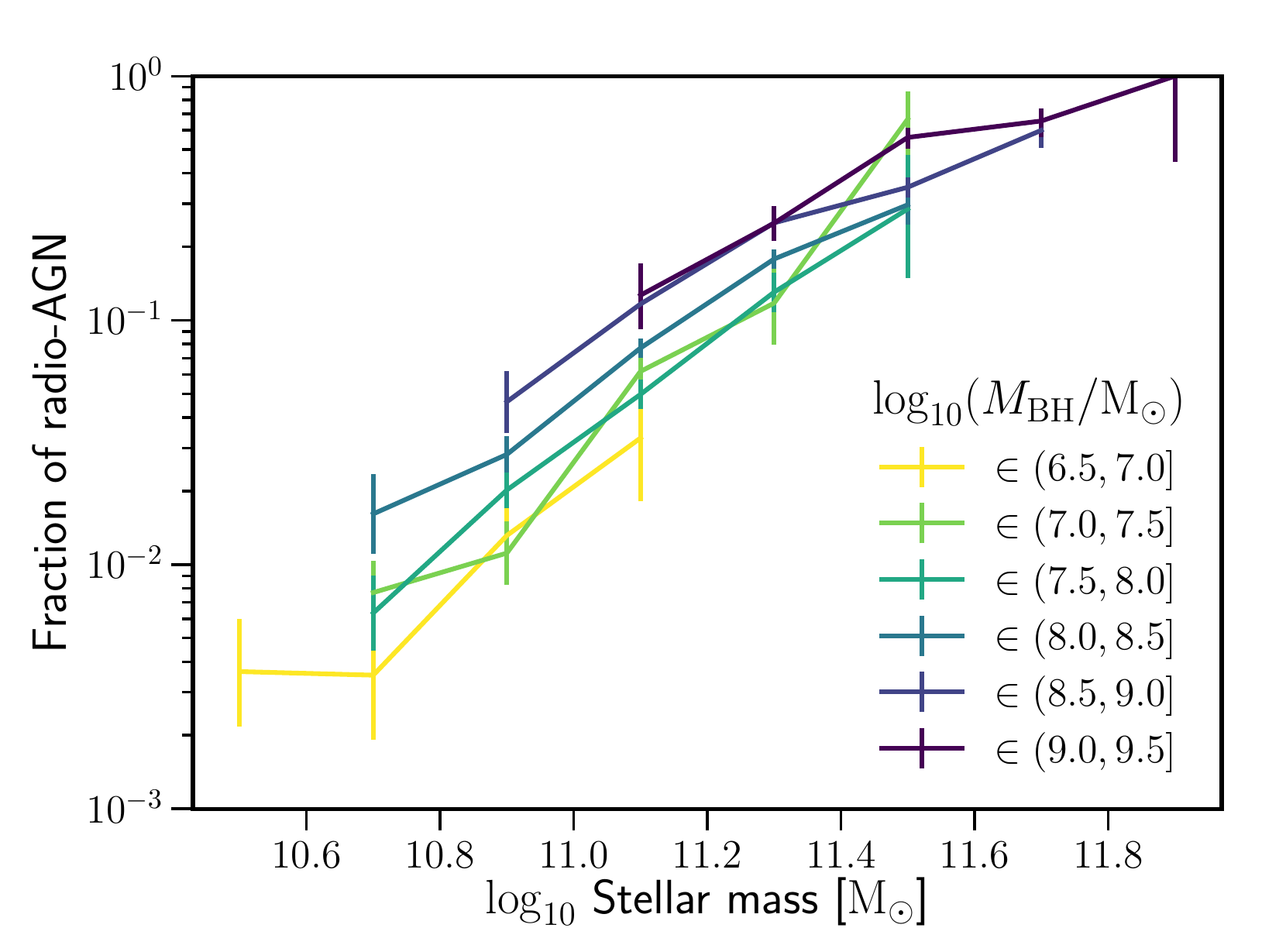}
   \caption{Left panel: Fraction of galaxies that host a radio AGN with 
$L_{\mathrm{150\,MHz}} \geq 10^{22.5}\,\mathrm{W\,Hz^{-1}}$, as a function of 
the black hole mass, for multiple stellar mass bins (shown in different 
colours). Right panel: The fraction of galaxies hosting radio-AGN with 
$L_{\mathrm{150\,MHz}} \geq 10^{22.5}\, \mathrm{W\,Hz^{-1}}$, as a function of 
the stellar mass, separated by their black hole mass (shown in different 
colours).}
   \label{fig:mass_mass_bh}
\end{figure*}

The stronger dependence of the radio-AGN fraction on stellar mass than on black 
hole mass found in Figs.~\ref{fig:fractions_stellar}~and~\ref{fig:fractions_bh} 
suggests that stellar mass is a more important factor in triggering radio-AGN 
activity. However, there is a strong correlation between black hole mass and 
stellar mass \citep[e.g.][]{Reines2015} that must be considered before reaching 
this conclusion. There is also a consideration that the black hole masses have 
larger uncertainties on their measurements because of the underlying scatter in 
the relationship between black hole mass and velocity dispersion from which they 
are estimated. In order to disentangle the effect of black hole mass and stellar 
mass, the fraction of radio AGN with respect to the black hole mass was computed 
in  bins of different stellar masses (left panel of Fig.~\ref{fig:mass_mass_bh}) 
and the fraction with respect to the stellar mass was computed in bins of 
different black hole masses (right panel of Fig.~\ref{fig:mass_mass_bh}) for 
galaxies with $L_{\mathrm{150\,MHz}} \geq 10^{22.5}\,\mathrm{W\,Hz^{-1}}$. This 
threshold in flux density maximises the number of galaxies available for the 
stratified study, which minimises the size of the error bars and scatter of the 
radio-AGN fractions, but consistent results are found for other radio luminosity 
limits. The error bars and fractions are computed as in 
Figs.~\ref{fig:fractions_stellar}~and~\ref{fig:fractions_bh}. From 
Fig.~\ref{fig:mass_mass_bh} it is clear that the fraction of radio AGN is mainly 
driven by the stellar mass. The black hole mass also has some residual effect 
but minimal in comparison with the effect of the stellar mass; indeed it is 
plausible that the remaining trend with black hole mass is simply a result of a 
residual correlation between black hole and stellar mass within the narrow 
stellar mass bin. The dominance of the stellar mass dependence points at an 
external trigger, rather than the properties of the black hole itself being the 
key driver. It is consistent with the properties of the gas that fuels the AGN 
being closely linked to the stellar mass of the galaxy 
\citep[e.g.][]{Sabater2013}, perhaps through the strong relationship between 
stellar mass and halo mass.

It is apparent in the left hand panels of both Fig.~\ref{fig:fractions_stellar} 
and Fig.~\ref{fig:fractions_bh} that there is a visible reduction in the slope 
of the $f_{\rm rad}$ versus $M_*$ relation from higher to lower luminosity 
limits. \citet{Janssen2012} split the radio-AGN population into high- and 
low-excitation sources, which correspond broadly to those fuelled at high 
Eddington rates by the accretion of cold gas in SFGs, and those fuelled at low 
rates by the accretion of gas from cooling hot haloes (e.g.\ 
\citealt{Hardcastle2007}; see discussion in \citealt{Heckman2014}). 
\citeauthor{Janssen2012} showed that the high-excitation population of radio AGN 
presents a more moderate dependency with the mass than low-excitation radio-AGN, 
and are more dominant at lower masses. The flattening in 
Fig.~\ref{fig:fractions_stellar} may well be related to this, as the lower radio 
luminosities typically probe lower stellar masses.


\section{Distribution of Eddington-scaled accretion rates}
\label{sec:accretion_rates}

The distribution of Eddington-scaled accretion rates of AGN has been previously 
studied by \citet{Kauffmann2009} and \citet{Best2012}. 
\citeauthor{Kauffmann2009} examined the accretion rates of emission-line 
selected AGN, deriving radiative AGN luminosities by scaling from the [O{\sc 
iii}]~5007 emission line. They found that red (quenched) galaxies followed a 
power-law distribution of Eddington-scaled accretion rates, rising towards lower 
accretion rates down to their lower observable limit of $L/L_{\mathrm{Edd}} \sim 
10^{-3}$. \citeauthor{Best2012} studied radio-selected AGN and derived 
Eddington-scaled luminosities by summing the radiative luminosity (scaled from 
[O{\sc iii}]~5007) and the jet mechanical luminosity (from the radio luminosity; 
see below). They also found that the distribution rose strongly down to 
$L/L_{\mathrm{Edd}} \sim 10^{-3}$ for the bulk of the radio population and that 
this was dominated by the jet mechanical luminosity except for the small 
population of strong emission line radio sources, which mostly occurred at high 
Eddington ratios \citep[see also discussion in][]{Heckman2014}. 

With the deeper LoTSS DR1 data it is possible to reach much lower limits in the 
Eddington ratio distribution. In this paper, this is only considered for the 
high stellar mass population (masses higher than $10^{11}$\,M$_{\odot}$). 
Consideration of just this population has two advantages: First, at these masses 
it was shown in Sec.~\ref{sec:fractions} that essentially all galaxies host a 
radio AGN, so the full Eddington rate distribution should be recoverable; 
second, at these masses the sample is expected to be dominated by the jet-mode 
AGN \citep[e.g.][]{Janssen2012}, meaning that a homogeneous population of 
sources is probed and that the Eddington ratio can be estimated using just the 
jet mechanical power.

The mechanical power of the radio jets in a given radio source would ideally be 
inferred directly from the properties of the radio source, by comparison with 
radio source evolution models \citep[see, for example,][]{Hardcastle2018}. 
However, such inference requires knowledge of the environments of the radio 
sources, which is not available for the current sample. Furthermore, radio 
source evolution models are generally most accurate for \citet{Fanaroff1974} 
Class II (FRII) sources, whereas at the low radio luminosities probed in this 
work most of the sources are either of FRI class or are compact. A common 
alternative approach for estimating the jet mechanical energy is therefore to 
scale directly from the radio luminosity. Such a conversion does not capture the 
intrinsic physical variations for individual sources; the radio luminosity of a 
radio source evolves throughout its lifetime even if the jet power remains 
constant\citep[ e.g.][]{Kaiser1997}. But for population-based analyses, such as 
that in this paper, the use of an average conversion factor can provide a 
suitably accurate approximation.

The jet mechanical power to radio luminosity conversion is commonly estimated 
from the cavities inflated by radio sources in the surrounding intergalactic or 
intracluster medium, as observed in X-rays. The total mechanical energy supplied 
by the radio jets is typically assumed to be 4$pV$, where $p$ is the pressure of 
the surrounding medium, $V$ the volume of the cavity, and the factor 4 comes 
from summing the work performed to inflate the cavities ($pV$) and the enthalpy 
of the relativistic plasma in the radio lobes (3$pV$). When combined with an 
estimate of the source age, for which the buoyancy timescale of the cavities is 
typically used \citep[e.g.][]{Churazov2001}, this allows the mechanical power of 
the jet to be estimated. This is found to correlate with the observed radio 
luminosity \citep{Birzan2004,Rafferty2006,Birzan2008,Cavagnolo2010}, although 
with significant scatter, as expected from the discussion above.

An alternative jet mechanical power to radio luminosity conversion was presented 
by \citet{Willott1999} based on the synchrotron properties of the radio source. 
For this, assumptions need to be made about the composition of the radio jet 
plasma \citep[see also][]{Croston2018}, and the cut-offs of the electron energy 
distribution. These are similar to the assumptions required for direct jet power 
inference for individual sources. 

\citet{Heckman2014} showed that the scaling relations determined from 
cavity-based methods and those of \citeauthor{Willott1999} provide broadly 
consistent estimates of the jet mechanical powers, at least at the 
moderate-to-high radio luminosities that dominate the energetic output of the 
AGN; this gives confidence that the assumptions made in each method are 
understood and broadly justified. They proposed the following 
population-averaged conversion between radio luminosity and jet 
mechanical power:$$P_{\rm mech,cav} = 2.8 \times 10^{37} \left(\frac{L_{\rm 
1.4\,GHz}}{10^{25}\,{\rm W\,Hz}^{-1}}\right)^{0.68}\,\mathrm{W}.$$

\noindent This expression is adopted in this work, modified to 150\,MHz 
assuming a spectral index of 0.7.

\begin{figure}   
   \centering
\includegraphics[width=0.86\linewidth]
{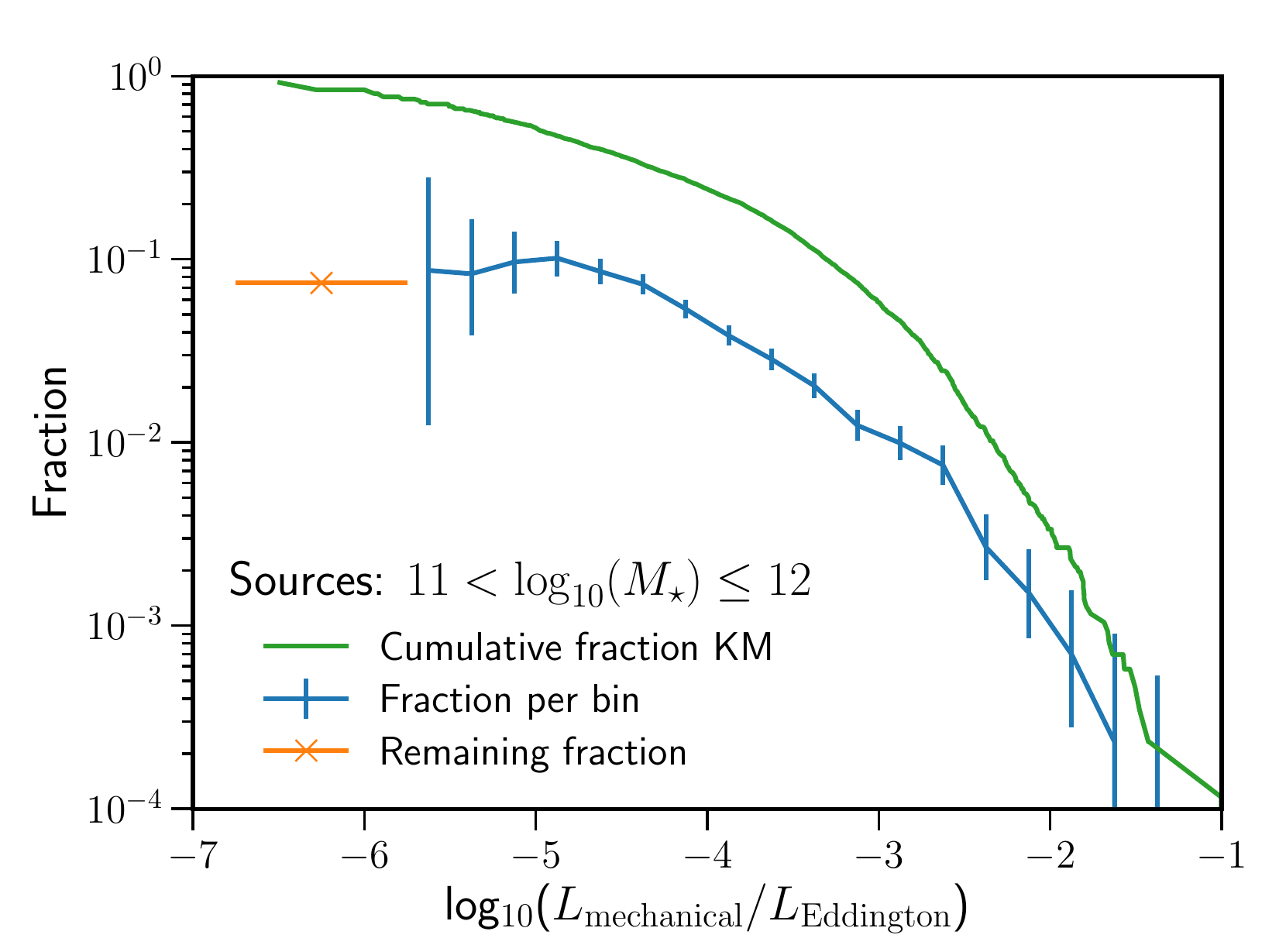}
   \caption{Distribution of Eddington-scaled accretion rate for radio AGN (solid 
blue line). The error bars correspond to the 95 per cent confidence interval. 
The cumulative distribution obtained from the Kaplan-Meier estimator (solid 
green line) running from higher towards lower accretion rates is shown. The 
orange cross represents the remaining fraction of galaxies spread over the 
following 4 bins. The distribution flattens and peaks at 
$L_{\mathrm{mech}}/L_{\mathrm{Edd}} \approx 10^{-5}$. A Kaplan-Meier estimator 
fit finds the median of the distribution at $L_{\mathrm{mech}}/L_{\mathrm{Edd}} 
= 10^{-4.98}$. We note that the conversion between radio luminosities and 
mechanical power is uncertain at low luminosities 
\citep[e.g.][]{Hardcastle2018}.}
   \label{fig:mech_over_ed}
\end{figure}

Using this, the distribution of Eddington-scaled accretion ratios for galaxies 
with masses between $10^{11}$\,M$_{\odot}$ and $10^{12}$\,M$_{\odot}$ is shown 
in Fig.~\ref{fig:mech_over_ed}.  For each bin in 
$L_{\mathrm{mech}}/L_{\mathrm{Edd}}$, this is calculated by considering the 
number of galaxies with detections in this range of accretion rates as compared 
to all the galaxies that could be detectable to that Eddington limit (with a 
detection above five times the rms noise level). The errors correspond to the 95 
per cent confidence interval using the Agresti-Coull binomial proportion 
confidence intervals. The accretion rate distribution increases from higher to 
lower accretion rates as previously determined, with this continuing down to 
$L_{\mathrm{mech}}/L_{\mathrm{Edd}} \approx 10^{-5}$ when the distribution 
flattens. A remaining small proportion of the galaxies have 
$L_{\mathrm{mech}}/L_{\mathrm{Edd}}$ below $10^{-5.75}$, indicated by the orange 
point on Fig.~\ref{fig:mech_over_ed} (which distributes these over the next four 
bins), indicating that the distribution must fall below 
$L_{\mathrm{mech}}/L_{\mathrm{Edd}} \sim 10^{-6}$.

The distribution of Eddington-scaled accretion rates was also checked using 
survival analysis. The Kaplan-Meier estimator \citep{Kaplan1958} was computed 
using the \texttt{Lifelines} package \citep{python_lifelines} in 
\texttt{Python}. Left-censoring was used, with detections marked as events and 
non-detections as limits using the $L_{\mathrm{mech}}/L_{\mathrm{Edd}}$ 
corresponding to five times the rms noise limit at the position. The results 
fully agree with the binned method shown before. The median of the distribution 
is found at $L_{\mathrm{mech}}/L_{\mathrm{Edd}} = 10^{-4.98}$; the mean of the 
distribution is at $L_{\mathrm{mech}}/L_{\mathrm{Edd}} = 10^{-3.53}$.

If all the galaxies belong to the same population the relative fractions can be 
directly linked to the time spent by the AGN at each accretion rate (duty 
cycle). It is interesting to note, therefore, that although the distribution 
peaks (i.e.\ galaxies spend most of their time) at relatively low accretion 
rates, the bulk of the time-averaged energy output for these galaxies is 
produced during the relatively short times corresponding to the higher 
Eddington-scaled accretion rates. The Kaplan-Meier distribution can be weighted 
by the energy output corresponding to each $L_{\mathrm{mech}}/L_{\mathrm{Edd}}$ 
and, integrating along this distribution, it is possible to find the average 
energy output corresponding to a given $L_{\mathrm{mech}}/L_{\mathrm{Edd}}$. In 
this way, it is found that 50 per cent of the energy output is emitted when 
$L_{\mathrm{mech}}/L_{\mathrm{Edd}} \ge -2.5$ ; the other 50 per cent is emitted 
at accretion rates below this value. Using the original distribution, the 
average time spent by a radio AGN at $L_{\mathrm{mech}}/L_{\mathrm{Edd}} \geq 
-2.5$ is only $\approx$ 1.7 per cent of the total. Hence, 50 per cent of the 
energy is released during the $\le 2$ per cent of the time spent at the highest 
accretion rates.

Considering that the distribution covers the full range of accretion rates, the 
total time-averaged mechanical energy output for high-mass radio AGN can be 
estimated. To do that, the mean accretion rate was computed from the mean 
Eddington-scaled value using the typical mass for the black holes in the 
selected stellar mass range: the median black hole mass of 
$10^{8.3}\,\mathrm{M_{\odot}}$ corresponds to $L_{\mathrm{Edd}} = 
10^{39.5}\,\mathrm{W}$, and this $L_{\mathrm{mech}}$ is expected to have an 
average value of $\approx 10^{36}\,\mathrm{W}$ ($10^{43}\,\mathrm{erg\,s^{-1}}$) 
through the lifetime and population of radio AGN. This result is broadly in line 
with the findings of \citet{Best2006}

It is interesting to compare this average AGN heating rate against the cooling 
radiation losses of the X-ray haloes surrounding these galaxies. 
\citet{Kim2015}, and references therein, consider scaling relations between 
X-ray luminosity and K-band luminosity. Considering the median stellar mass of 
the galaxies in the mass range selected above ($10^{11.21}\, 
\mathrm{M_{\odot}}$) and assuming that for these old ellipticals the K-band 
luminosity can be used as a proxy of the stellar mass (with a mass-to-light 
ratio of order unity), this AGN heating rate is significantly above the X-ray 
luminosity expected for individual ellipticals. Instead, it is more similar to 
the values found by \citeauthor{Kim2015} for the central galaxies of groups and 
clusters, and of the same order of magnitude of that associated with the 
inflation of bubbles/cavities in clusters \citep[e.g.][]{McNamara2007, 
Diehl2008}. This is consistent with recurrent radio-AGN activity in these 
high-mass galaxies providing sufficient energy to control cooling in their 
surrounding groups and clusters and providing the feedback that maintains the 
host galaxies as quenched, as expected \citep[see also][]{Hardcastle2018}.

\section{Summary and conclusions}
\label{sec:conclusions}

The LoTSS DR1 and the SDSS main spectral galaxy sample have been combined to 
study the prevalence of radio AGN in the local Universe ($0.01 \leq z \leq 
0.3$). There are 33504 SDSS galaxies in the LoTSS area and 32 per cent of these 
are detected by LoTSS. The main results of this study are as follows:

\begin{itemize}
\item The method to classify and separate radio AGN from SFGs has been adapted 
to the 150 MHz frequency of LoTSS and improved by both the inclusion of WISE 
colours and the calibration of diagnostic lines against the well-studied sample 
of \citet{Gurkan2018}. A total of 2121 radio AGN were identified.

\item The distribution of 150\,MHz to 1.4\,GHz spectral indices was obtained. A 
median spectral index of 0.63, in line with the canonical value of 0.7, is found 
at high flux densities ($S_{\rm 1.4\,GHz} > 20$\,mJy). The median spectral index 
decreases at lower flux densities, but simulations indicate that this may be 
entirely driven by selection effects whereby steeper spectrum sources are missed 
in the shallower high-frequency data sets.

\item The local radio luminosity function at 150\,MHz has been derived for radio 
AGN and SFGs separately down to lower luminosities than has previously been 
possible. The radio-AGN luminosity function agrees well with previous 
determinations of the luminosity function at 1.4\,GHz, which gives confidence in 
the robustness of the AGN/SF separation method.

\item The prevalence of radio AGN, as selected at 150\,MHz, has been studied as 
a function of the stellar and black hole masses. As previously seen at high 
luminosity at higher frequencies, the fraction of galaxies hosting radio AGN 
rises strongly with mass. This relation is seen to be flatter at lower 
luminosities and lower stellar masses.

\item Stellar mass is a stronger driver of the fraction of AGN than the black 
hole mass. Indeed, once the correlation between black hole and stellar mass is 
accounted for, the radio-AGN fraction still rises strongly with stellar mass (at 
fixed black hole mass) but shows little or no dependence on black hole mass (at 
fixed stellar mass). This is indicative of the stellar mass being better linked 
to the properties of the gas that fuels the AGN and drives the activity.

\item Remarkably, 100 per cent of galaxies with  masses higher than 
$10^{11}\,\mathrm{M_{\odot}}$ host radio AGN when limits below or equal to 
$L_{\mathrm{150\,MHz}} \geq 10^{21}\,\mathrm{W\,Hz^{-1}}$ are considered. This 
suggests that the most massive galaxies are always switched on, even if it is at 
relatively low radio powers.

\item The full distribution of Eddington-scaled accretion rates was derived for 
the most massive galaxies (stellar masses between $10^{11}$\,M$_{\odot}$ and 
$10^{12}$\,M$_{\odot}$). This is shown to rise with decreasing Eddington 
fraction down to $L_{\mathrm{mech}}/L_{\mathrm{Edd}} \sim 10^{-5}$ and to drop 
sharply below $L_{\mathrm{mech}}/L_{\mathrm{Edd}} \sim 10^{-6}$. However, the 
bulk of the energy output is produced during the relatively short times that AGN 
are powering jets close to their maximum powers.
\end{itemize}

The use of LoTSS has permitted the exploration of a wider parameter space than 
previous studies. The results are very much in line with previous work, but the 
extension to lower luminosities has greatly extended the scope of that work. The 
discovery that the most massive galaxies are always on as a radio source at the 
luminosity levels that LoTSS reaches and that stellar mass appears to be a more 
important driver of activity than black hole mass, both indicate that the 
radio-AGN activity is controlled by the fuel supply to the radio AGN, which 
connects more closely to the stellar mass. This fits the popular picture whereby 
this fuel is connected to cooling of hot gas within the dark matter halo, and 
the radio-AGN activity is instrumental in maintaining the host galaxies as old, 
red, and dead.

Considering that the DR1 of LoTSS covers only 2 per cent of the final LoTSS 
area, an improvement of at least one order of magnitude in the volume of LoTSS 
data can be expected in the near future (next two to three years). These larger 
samples will allow a more detailed exploration of additional aspects of the 
triggering of radio AGN. In particular, it will be very interesting to directly 
test the picture above by separating the radio-AGN luminosity distribution by 
additional parameters such as environment \citep[see][]{Sabater2013,Sabater2015} 
and mass to separate the roles of these two critical parameters. Furthermore, 
the forthcoming availability of extensive spectroscopic data for LoTSS sources 
out to higher redshifts, through the WEAVE-LOFAR survey \citep[][]{Smith2016}, 
will allow cosmic evolution to be examined. These topics will constitute the 
focus of future detailed studies. 

\begin{acknowledgements}
We would like to thank the anonymous referee for useful comments that 
improved this paper.

This paper is based (in part) on data obtained with the International LOFAR 
Telescope (ILT) under project codes LC2\_038 and LC3\_008. LOFAR 
\citep{vanHaarlem2013} is the LOw Frequency ARray designed and constructed by 
ASTRON. It has observing, data processing, and data storage facilities in 
several countries, which are owned by various parties (each with their own 
funding sources) and are collectively operated by the ILT foundation 
under 
a joint scientific policy. The ILT resources have benefited from the following 
recent major funding sources: CNRS-INSU, Observatoire de Paris and Université 
d'Orléans, France; BMBF, MIWF-NRW, MPG, Germany; Science Foundation Ireland 
(SFI), Department of Business, Enterprise and Innovation (DBEI), Ireland; NWO, 
The Netherlands; The Science and Technology Facilities Council, UK.

Funding for the SDSS and SDSS-II has been provided by the Alfred P. Sloan
Foundation, the Participating Institutions, the National Science Foundation, the
U.S. Department of Energy, the National Aeronautics and Space Administration,
the Japanese Monbukagakusho, the Max Planck Society, and the Higher Education
Funding Council for England. The SDSS Web Site is http://www.sdss.org/.
The SDSS is managed by the Astrophysical Research Consortium for the
Participating Institutions. The Participating Institutions are the American
Museum of Natural History, Astrophysical Institute Potsdam, University of Basel,
University of Cambridge, Case Western Reserve University, University of Chicago,
Drexel University, Fermilab, the Institute for Advanced Study, the Japan
Participation Group, Johns Hopkins University, the Joint Institute for Nuclear
Astrophysics, the Kavli Institute for Particle Astrophysics and Cosmology, the
Korean Scientist Group, the Chinese Academy of Sciences (LAMOST), Los Alamos
National Laboratory, the Max-Planck-Institute for Astronomy (MPIA), the
Max-Planck-Institute for Astrophysics (MPA), New Mexico State University, Ohio
State University, University of Pittsburgh, University of Portsmouth, Princeton
University, the United States Naval Observatory, and the University of
Washington.

This work was carried out in part on the Dutch national e-infrastructure with 
the support of SURF Cooperative through e-infra grants 160022 \& 160152.

PNB, JS, and RKC are grateful for support from the UK Science and Technology 
Facilities Council (STFC) via grant ST/M001229/1 and ST/R000972/1.

KJD and AS acknowledge support from the European Research Council under the 
European Unions Seventh Framework Programme (FP/2007-2013)/ERC Advanced Grant 
NEWCLUSTERS-321271

MJH and WLW acknowledge support from the UK Science and Technology Facilities 
Council 
ST/M001008/1.

JHC acknowledges support from the STFC under grants ST/R00109X/1 and 
ST/R000794/1.

APM would like to acknowledge the support from the NWO/DOME/IBM programme "Big 
Bang Big Data: Innovating ICT as a Driver For Astronomy", project \#628.002.001.

GG acknowledges the CSIRO OCE Postdoctoral Fellowship.

F.d.G. is supported by the VENI research programme with project number 1808, 
which is financed by the Netherlands Organisation for Scientific Research (NWO).

LKM acknowledges the support of the Oxford Hintze Centre for Astrophysical 
Surveys, which is funded through generous support from the Hintze Family 
Charitable Foundation. This publication arises from research partly funded by 
the John Fell Oxford University Press (OUP) Research Fund.

IP acknowledges support from INAF under PRIN SKA/CTA ‘FORECaST’

SM acknowledges funding through the Irish Research Council New Foundations 
scheme and the Irish Research Council Postgraduate Scholarship scheme.

This research made use of \textsc{Astropy}, a community-developed core Python
package for
Astronomy (Astropy Collaboration, \citeyear{astropy}, \citeyear{astropy2}); 
\textsc{Ipython}
\citep{IPython}; \textsc{matplotlib} \citep{matplotlib}; \textsc{numpy}
\citep{numpy}; \textsc{pandas} \citep{pandas}; \textsc{scipy}
\citep{scipy}, \textsc{TOPCAT} \citep{TOPCAT}, and \textsc{KERN} suite 
\citep{Kern}.

\end{acknowledgements}

\bibliographystyle{aa}
\bibliography{databasesol}

\label{lastpage}
\end{document}